\documentclass[12pt]{article}
\usepackage{graphicx}
\usepackage{mathrsfs}
\usepackage{dsfont}
\usepackage{graphicx}
\usepackage{units}
\usepackage{amsmath, amsthm, amssymb, amsfonts, enumerate}
\usepackage{natbib}
\usepackage{color}
\usepackage{caption}
\usepackage{subcaption}
\usepackage{setspace}
\usepackage{mdframed}
\usepackage{hyperref}
\usepackage{float}
\usepackage{multirow} % Add this to the preamble for the multirow feature
\usepackage[normalem]{ulem}
\usepackage{placeins}
\usepackage{tikz}
\usepackage{xcolor}
\usepackage{titlesec}
\usetikzlibrary{calc}
\usepackage[T1]{fontenc}
\usepackage{titling}

\definecolor{blue}{rgb}{0.0, 0.0, 1}
\definecolor{green2}{rgb}{0.0, 0.52, 0.24}
\definecolor{cadmiumgreen}{rgb}{0.0, 0.42, 0.24}
\definecolor{camouflagegreen}{rgb}{0.47, 0.53, 0.42}
\definecolor{darkolivegreen}{rgb}{0.33, 0.42, 0.18}
\definecolor{darkpastelgreen}{rgb}{0.01, 0.75, 0.24}
\definecolor{darkspringgreen}{rgb}{0.09, 0.45, 0.27}
\definecolor{darkspringgreen}{rgb}{0.09, 0.45, 0.27}

\usepackage{comment} %commenting bubbles package.
\usepackage[textwidth=30mm,disable]{todonotes}

\def\o{{\omega}}

\newcommand{\heading}[1]{\noindent \textbf{#1}}

\long\def\ignore#1{}
\def\a{\alpha}
\def\b{\beta}
\def\l{\ell}

\def\1{\textbf{1}}

\usepackage[letterpaper,margin=1.25in]{geometry}
\usepackage{setspace}
\newtheorem{theorem}{Theorem}

\newtheorem{proposition}{Proposition}
\newtheorem{corollary}{Corollary}
\newtheorem{example}{Example}
\newtheorem{definition}{Definition}

\newtheorem{remark}{Remark}
\newtheorem{lemma}{Lemma}
\newtheorem{obs}{Observation}

\definecolor{plusfill}{RGB}{232,242,255}
\definecolor{plusdraw}{RGB}{31,78,121}
\definecolor{minusfill}{RGB}{253,234,234}
\definecolor{minusdraw}{RGB}{178,34,34}
\definecolor{arrowgray}{RGB}{100,100,95}

\tikzset{
  pluscell/.style={
    draw=plusdraw,
    fill=plusfill,
    rounded corners=8pt,
    minimum width=3.85cm,
    minimum height=1.9cm,
    inner sep=9pt,
    align=left,
    text=plusdraw
  },
  minuscell/.style={
    draw=minusdraw,
    fill=minusfill,
    rounded corners=8pt,
    minimum width=3.85cm,
    minimum height=1.9cm,
    inner sep=9pt,
    align=left,
    text=minusdraw
  },
  loopplus/.style={
    draw=plusdraw,
    fill=plusfill,
    rounded corners=8pt,
    minimum width=3.8cm,
    minimum height=1.45cm,
    inner sep=7pt,
    align=center,
    text=plusdraw
  },
  loopminus/.style={
    draw=minusdraw,
    fill=minusfill,
    rounded corners=8pt,
    minimum width=3.8cm,
    minimum height=1.45cm,
    inner sep=7pt,
    align=center,
    text=minusdraw
  },
  looparrow/.style={
    ->,
    thick,
    draw=arrowgray,
    >=Stealth
  }
}

\newcommand{\twostatecell}[5]{%
\begin{minipage}{3.25cm}
\(\boldsymbol{#1}\)\\[5pt]
\(#2\)\hfill \(#3\)\\[4pt]
\(#4\)\hfill \(#5\)
\end{minipage}
}

\begin{document}

\title{Public Information Generators: Common Knowledge and Information Loops\thanks{For their valuable comments, the authors wish to thank participants of the $37^{\rm th}$ Stony Brook Game Theory Conference, the Paris Workshop on Game Theory and Language: Half a Century of ``Agreeing to Disagree'', the University of Toronto Economic Theory Workshop, the Durham University Economics Seminar, the Adam Smith Business School Micro Theory Seminar of Glasgow University, INSEAD EPS seminar, The School of Economics seminar at the University of Edinburgh, Western University Economics seminar, the Rationality Center Game Theory Seminar, the Technion Game Theory Seminar, the Bar-Ilan University Theoretical Economics Seminar, the Tel Aviv University Game Theory Seminar,  the Bar-Ilan University Management Seminar and the BGU Economics seminar.
Lagziel acknowledges the support of the Israel Science Foundation, Grant \#2074/23. Lehrer acknowledges the support of the Deutsche Forschungsgemeinschaft (DFG, German Research Foundation), Project Number 461570745. Wang acknowledges the support of the National Natural Science Foundation of China \#72303161. }}
\author{David Lagziel\thanks{Department of Economics, Ben-Gurion University of the Negev, Beer-Sheba 8410501, Israel.  E-mail: \textsf{davidlag@bgu.ac.il}.} 
\and
Ehud Lehrer\thanks{Economics Department, Durham University, Durham DH1 3LB, UK.  E-mail: \textsf{ehud.m.lehrer@durham.ac.uk}.}
\and
Tao Wang\thanks{International School of Economics and Management, Capital University of Economics and Business, Beijing 100070, China.  E-mail: \textsf{tao.wang.nau@hotmail.com}.}  }
\maketitle

\thispagestyle{empty}

\begin{abstract}
\singlespacing{
We analyze incomplete-information games where an oracle publicly shares information with privately informed players. One oracle dominates another if, for every experiment of the latter, it can choose an experiment that supports, in every game, every equilibrium outcome distribution induced by the other. We fully characterize equivalence (mutual dominance) and identify the information-loop obstruction governing one-sided dominance, obtaining necessity in general and sufficiency under separated-loop structures. The analysis highlights the role of common-knowledge components and develops a theory of information loops, thereby extending the seminal work of \cite{Blackwell1951} to strategic environments and \cite{Aumann1976}'s theory of common knowledge.}
\end{abstract}



\newpage
\setcounter{page}{1}

%--------------------------------------------------------------------------
%--------------------------------------------------------------------------
\section{Introduction} \label{Section - Intro}
%--------------------------------------------------------------------------
%--------------------------------------------------------------------------

% The motivation for the paper 

Public information in strategic environments does more than reveal facts: it changes how privately informed agents reason about one another. As a result, signals released by intermediaries may alter equilibrium behavior even when the intermediaries themselves have only partial information. This introduces a fundamental comparative problem: characterizing when one information intermediary is universally more informative than another in strategic interactions. The question is central in settings ranging from financial markets to online platforms, where public information shapes strategic behavior by selectively revealing information to heterogeneously informed agents.

% High level perspective on the problem

This paper examines incomplete-information games where players are partially informed, both privately and publicly, about the realized state.  The private information is provided to every player by their specific partition, and the public information is disclosed by an external source, namely, an \emph{oracle}.

% What is an oracle and what it does?

Formally, an oracle holds a partition of the state space and communicates through a \emph{signaling function} that is measurable with respect to the oracle's partition.  Any such signaling function is a Blackwell experiment (see \citealp{Blackwell1951}), so an oracle can be viewed as \emph{a generator of experiments} compatible with its information.  Each experiment, combined with the players’ private partitions, induces a ``guided game" with its own set of equilibria. Notably, the oracle need not know what common knowledge among the players is.

%The definition of dominance among oracles

The primary objective is to compare two oracles in terms of their ability to induce equilibria across all games.  Oracle 1 is said to \emph{dominate} Oracle 2 if, for every experiment available to Oracle 2, Oracle 1 can generate an experiment, such that in every game, the set of equilibrium distributions over state-action profile pairs contains the one generated by Oracle 2.

%3 remarks on the definition

Throughout the paper, we hold the players' private information fixed, isolating the variation strictly to the oracles' experiments. Doing so, we adopt an implementability notion of dominance requiring Oracle~1 to support every equilibrium outcome distribution induced by Oracle~2. This comparison does not assign an objective to the oracle, and contrasts with alternative notions that require the comparison to hold \emph{for every} configuration of the players' private information. Further discussion of these modeling assumptions and the challenges they create is given in Section~\ref{Section - Model}.

%Main contribution and loop geometry

Dominance has two complementary components: a local component and a global one. The local part is governed by the players' \emph{common knowledge components} (CKCs), the minimal commonly known events (see \citealp{Aumann1976}). The global part is governed by a new structure, referred to as \emph{information loops}: closed paths that link different CKCs through atoms of the oracle's partition. To see why this structure is needed, consider three CKCs \(C_j=\{\omega_j,\overline\omega_j\}\), \(j=1,2,3\), under a uniform prior. Let Oracle~1's partition be $F_1=\{\{\overline{\omega}_1,\omega_2\},\{\overline{\omega}_2,\omega_3\},\{\overline{\omega}_3,\omega_1\}\}$, as in Figure~\ref{fig: F1 irreducible loop}, and let Oracle~2's partition be \(F_2=\{\{\omega_1,\omega_2,\omega_3\},\{\overline\omega_1,\overline\omega_2 ,\overline\omega_3\}\}\). Inside each CKC, both oracles separate the two states, so they are equally informative locally. Yet Oracle~2 can send a signal \(s\) with probability \(1/3\) at each \(\omega_j\) and \(2/3\) at each \(\overline\omega_j\), producing likelihood ratio \(1/2\) in every CKC. Oracle~1 cannot reproduce this: every positive-probability mimicking signal of Oracle~1 would need a likelihood ratio in $\{1/2,2\}$ in each CKC, while $F_1$-measurability requires the product of the three ratios to be one. Thus, local informativeness is not enough, as likelihood ratios inside different CKCs must be globally compatible with a single oracle-measurable public signal.

\begin{figure}[H]
        \centering
        \begin{tikzpicture}[scale=0.8]

        % Draw the rectangles with labels
        \draw (1.5,0.5) -- (4.5,0.5) -- (4.5,1.5) -- (1.5,1.5) -- cycle;
        \node at (2,1) {$\o_1$};
        \node at (4,1) {$\overline{\o}_1$};

        \draw (4.5,-1) -- (5.5,-1) -- (5.5,-3.5) -- (4.5,-3.5) -- cycle;
        \node at (5,-1.5) {$\o_2$};
        \node at (5,-3) {$\overline{\o}_2$};

        \draw (0.5,-1) -- (1.5,-1) -- (1.5,-3.5) -- (0.5,-3.5) -- cycle;
        \node at (1,-3) {$\o_3$};
        \node at (1,-1.5) {$\overline{\o}_3$};

        % Correct the ellipses
        \draw[blue, thick, rotate around={-65:(4.45,-0.25)}] (4.45,-0.25) ellipse (2.0cm and 0.7cm); % Blue ellipse
        \node[blue] at (4.4,-0.2) {$F_1(\omega_2)$};
        \draw[red, thick, rotate around={0:(3,-3)}] (3,-3) ellipse (2.4cm and 0.7cm); % Red ellipse
        \node[red] at (3,-3) {$F_1(\omega_3)$};
        \draw[teal, thick, rotate around={65:(1.5,-0.25)}] (1.5,-0.25) ellipse (2.0cm and 0.7cm); % Green ellipse
        \node[teal] at (1.5,-0.2) {$F_1(\omega_1)$};

        % Add labels
        \node at (3,1.8) {$C_1$};
        \node at (6,-2.5) {$C_2$};
        \node at (-0.1,-2.5) {$C_3$};

        \end{tikzpicture}

     \caption{ \footnotesize An illustration of an $F_1$-loop. The loop intersects three CKCs ($C_1, C_2, C_3$), each containing two states, via three distinct atoms of Oracle 1's partition ($F_1$).}
    \label{fig: F1 irreducible loop}
\end{figure}

Information loops capture this additional layer of structure. They arise from the interaction between the players' common-knowledge structure and the external oracle's partition, and they determine whether local mimicking constructions inside CKCs can be glued into a single oracle-measurable experiment. As a consequence, refinement within every CKC is not sufficient for dominance: even a locally more informative oracle may face cyclic compatibility constraints across CKCs.

Before developing the loop analysis, Section~\ref{Section:characterization using oracles' strategies} records a posterior characterization of dominance. Proposition~\ref{Prop - Charact in terms of signaling funct} shows that Oracle~1 dominates Oracle~2 if and only if every experiment of Oracle~2 can be matched by an experiment of Oracle~1 in terms of the distribution of posterior beliefs generated within each CKC. This result is used throughout the analysis to translate strategic dominance into posterior mimicry.

Our main analysis begins with the study of loop geometry. We identify several essential properties of information loops, the central one being the existence of an \emph{order-preserving cover}. An $F_1$-loop admits an order-preserving $F_2$-cover if its pairs can be partitioned into $F_2$-sub-loops and a set of pairs that $F_2$ does not distinguish, with each sub-loop preserving the cyclic order inherited from the original $F_1$-loop. This property, along with several others studied in Section~\ref{Section: Information loops}, underlies our two main results.

The first main result, Theorem~\ref{Theorem: Equivalent oracles}, characterizes \emph{equivalent} oracles, defined through mutual dominance. The equivalence problem is the cleanest benchmark: mutual dominance forces both oracles to agree locally within every CKC and globally on their primitive loop structure. Specifically, Oracles~1 and~2, with partitions $F_1$ and $F_2$ respectively, are equivalent if and only if $F_1$ and $F_2$ refine each other within every CKC and every irreducible\footnote{An $F_i$-loop is irreducible if its states do not comprise a shorter $F_i$-sub-loop.} $F_i$-loop admits an order-preserving $F_{-i}$-cover, for $i=1,2$.

We then turn to one-sided dominance, Theorem~\ref{Theorem: stochastic loop cover dominance}, our second main result and a loop-cover theorem. If Oracle~1 dominates Oracle~2, then $F_1$ refines $F_2$ inside every CKC, every irreducible $F_1$-loop is $F_2$-covered, and every unique cover of an irreducible loop must preserve the cyclic order of the original loop. Conversely, local refinement together with order-preserving covers of irreducible loops implies dominance under separated-loop structures. Thus, one-sided dominance is governed by local refinement and by the global compatibility of cyclic likelihood-ratio constraints.\footnote{Notably, by confining the oracles to deterministic signaling functions, \cite{Lagziel2026a} provide a complete characterization using a simultaneous posterior-matching condition.}

Finally, these results immediately yield several benchmark cases. For example, if the state space consists of a single CKC, or if there are no $F_1$-loops, the global loop condition is vacuous, and dominance collapses to refinement within every CKC (Corollary~\ref{Corollary: stochastic benchmark cases}).

\subsection{Who generates public information?}
An oracle is any institution or intermediary that makes public statements on the basis of potentially incomplete information, that might not be identical to the information held by the players. The oracle's partition represents what this intermediary can observe, verify, or certify, and its experiment represents how that evidence is translated into public information. The object being compared is therefore neither a single signal nor a specific experiment, but a technology for generating public information.

Public disclosure by regulators is a leading example. In financial supervision, investors, creditors, and analysts hold heterogeneous private information about financial institutions and market conditions. A central bank or supervisory authority may observe stress-test data or confidential balance-sheet information, but must still choose how to disclose it. The supervisor's evidence determines what its experiment can condition on, while the disclosure rule determines the experiment itself. Comparing two transparency regimes is therefore a comparison between two experiment-generating technologies, not merely between two realized announcements.

A related interpretation arises in public mediation. In commercial disputes, 
regulatory settlements, or political negotiations, the parties hold private information 
that shapes the scope for agreement. A mediator may observe evidence that cuts 
across these private fragments and then issue a public statement that changes the parties' 
beliefs about the state and about one another. Our framework ranks such 
mediation technologies without assigning a payoff function to the mediator or imposing 
an equilibrium-selection rule. This differs from standard persuasion, where the 
sender's objective is specified, and is complementary to work on coordination under 
strategic uncertainty, where a designer jointly uses incentives and information to 
guarantee desirable outcomes (see \citealp{Halac2022}).

By contrast, we study public-information technologies in which the intermediary neither observes nor elicits players' private information. Allowing multidirectional communication and player-specific private recommendations would enable richer forms of coordination. Our model instead restricts the intermediary to a public experiment measurable with respect to its own partition. Public labels may still coordinate equilibrium selection, but only through a signal observed by all
players. This resembles a firm making a public market announcement without first eliciting investors' heterogeneous private information. Accordingly, the model isolates how public information interacts with fixed common-knowledge components and separates public disclosure from elicitation-based mechanism design.

A mediation-game application is developed in Section~\ref{Section - A simple example},  and a fact-checking example is given in Appendix~\ref{Appendix: polarization example}. These applications share a structure that is not captured by standard concavification tools.\footnote{\cite{Lagziel2025f} show that the feasible set of joint posterior beliefs is typically non-convex given a non-binary state space, when both the oracle and the players possess private and partial information.} The intermediary cannot choose arbitrary Bayes-plausible distributions over joint posteriors, but can only choose experiments measurable with respect to its own information, and the effect of any such experiment depends on how it interacts with the players' private partitions. Oracle dominance addresses this directly by asking whether one intermediary can replicate the joint posterior profiles, and hence the equilibrium outcome distributions, that another can generate. Two structural objects govern the comparison: the CKCs of the players' partitions, which determine local mimicking, and the information loops linking these components, which determine global compatibility. The theory then ranks the generators of public information, but without solving a separate information-design problem for each possible objective. 

\subsection{Information endowments}

Our framework assumes that the initial information endowments of all agents, oracles and players, are given by partitions of the state space, rather than by stochastically generated information structures. This restriction applies only to their initial information: oracles may employ stochastic experiments to transmit information to the players.

This structural boundary is methodologically deliberate. Evaluating the comparative value of oracles in a multi-player environment introduces substantial combinatorial and strategic complexity. Indeed, characterizing the feasible posterior regions, the limits of information transmission, and the resulting strategic implications is already a highly non-trivial mathematical task when initial information is partitioned, even before introducing general type spaces.

Consequently, focusing on partitions allows us to isolate the fundamental mechanisms of the model without the confounding analytical tractability issues introduced by stochastic noise. This framework serves as the necessary foundation for this line of research. The formal analysis established here provides the benchmark structures and tools required to extend the model to fully stochastic information environments in subsequent studies.  This baseline partition structure also clarifies the boundaries of the signaling mechanism itself.

\subsection{Relation to literature}
\cite{Blackwell1951,Blackwell1953} provides the classic comparison of experiments for a single decision maker: one information structure dominates another if it yields weakly higher expected utility in every decision problem. We extend this framework along two dimensions. First, we move from decision problems to incomplete‑information games, where the objects being compared are equilibria of ``guided games'' induced by an oracle's signals. Second, instead of a fixed experiment, an oracle is a generator of experiments compatible with its partition. Our dominance notion is therefore about the ability of one oracle to replicate the equilibrium outcome distributions that another oracle can induce, rather than optimizing a particular decision maker’s payoff.

\cite{Brooks2024} strengthen Blackwell by requiring robustness to arbitrary auxiliary signals and decision problems and characterize \emph{strong Blackwell dominance} between two signals. In contrast, we fix the players' private information structures and the prior, and compare oracles that can implement any experiment measurable with respect to their partitions. Our dominance relation is specific to a given configuration of players' information and to strategic interaction, not to a universal set of decision problems.

A related literature compares information structures in games and establishes partial orders. \cite{Peski2008} analyzed zero-sum games, offering an analogous result to Blackwell's by characterizing when one information structure is more advantageous for the maximizer.
\cite{Lehrer2010,Lehrer2013} analyze signaling and mediation in common‑interest games and show how variants of Blackwell garbling characterize outcome equivalence. Likewise, \cite{Bergemann2016} characterize dominance between two information structures through the concept of individual sufficiency, an extension of Blackwell’s notion of garbling to $n$-player games. A common feature of these papers is that they compare \emph{fixed} information structures (typically private signals) and use versions of Blackwell’s garbling to capture dominance. We instead hold players’ private information fixed and compare oracles that provide  \emph{public signals}, subject to an oracle-measurability constraint (the oracle cannot condition on players’ private signals). For this reason, oracle dominance is not meant to coincide with, or simply specialize, \cite{Bergemann2016} to public signals: even under a “public-signal-only” restriction, their order is a pairwise comparison of two fixed information structures, whereas ours is a comparison over the menus of feasible public experiments generated by different oracle partitions. Dominance in our setting is driven by the implementable joint posterior beliefs, constrained by CKCs and information loops.

Another strand studies mediators in incomplete-information games who correlate players' actions through private recommendations, often without adding information about the realized state; see \cite{Forges1993} and \cite{Gossner2000} among others. Our oracles instead communicate through a common public signal and cannot issue player-specific recommendations. Thus, oracles can affect outcomes through both belief changes and public coordination. The distinction from standard mediation is that this coordination must be generated by a single public experiment measurable with respect to the oracle's partition, rather than through privately correlated recommendations. Consequently, oracle dominance cannot be reduced to the richness of the correlated-equilibrium correspondence.

Our departure from existing dominance notions (see, e.g., \citealp{Gossner2000} and \citealp{Bergemann2016}) lies in restricting information provision to \emph{public disclosure}, given players' private information and the Oracle's information. This restriction creates global measurability and consistency constraints across CKCs. If players' private information is trivial, or if the comparison is required to hold uniformly over all possible player-information structures, the loop constraints disappear and dominance reduces to partition refinement. Allowing private recommendations instead defines a different mediation problem and does not, by itself, remove the oracle-measurability constraint. The conceptual contribution of this paper is to isolate and clarify the role of public disclosure in shaping the interaction between fixed players' private information and the Oracle's information, and to show how this interaction fundamentally alters the set of attainable equilibrium outcomes.

Our use of CKCs is rooted in the epistemic foundations of games.  \cite{Aumann1976} defines common knowledge and the induced partition into CKCs. \citet{Monderer1989}, \citet{Mertens1985}, \citet{Brandenburger1993}, and \cite{Battigalli1999} clarify how hierarchies of beliefs and type spaces encode such information. Our model builds on these studies by fixing the partition structures while varying only the oracle's public experiment.  The novel constraints we study arise from \emph{global} measurability across CKCs (via loops), not from additional complexity in private belief hierarchies.  Information loops then formalize how public measurability links different CKCs and constrains the set of posterior profiles that an oracle can generate.

The dominance problem, and in particular the analysis of order-preserving covers, can be partially translated into the study of Eulerian directed multigraphs. In the unique-cover case, the associated \(F_2\)-atom multigraph has a unique directed-cycle partition and, by \cite{Cooper2025}, is a bridgeless cactus; in particular, distinct covering cycles intersect in at most one \(F_2\)-atom. We are not aware of a general characterization of the order-preservation problem beyond such benchmark regimes.

\subsection{The structure of the paper}
Section~\ref{Section - A simple example} presents the motivating example. Section~\ref{Section - Model} introduces the model, dominance, and posterior mimicry. Section~\ref{Section: Information loops} develops local refinement and loop geometry. Sections~\ref{Section - Equivalent oracles} and~\ref{Section - necessary and sufficient conditions for dominance} present the equivalence and one-sided dominance results. Section~\ref{Section - Partial ordering of oracles} collects the benchmark cases, and Section~\ref{Section - Conclusion} concludes.

\section{A mediation game: motivating example} \label{Section - A simple example}

The following example shows that two oracles carrying identical information within every CKC can nonetheless support markedly different sets of equilibrium outcomes. Consider a game with two players, indexed by $i=1,2$, each with two actions: either \emph{attack} (A) or \emph{compromise} ($C$).  The set of states is $\Omega=\{\omega_1,\omega_2,\omega_3,\omega_4\}$, endowed with a uniform common prior.  The players hold asymmetric information: player~1’s partition is $\Pi_1=\{\{\omega_1,\omega_2\},\{\omega_3\},\{\omega_4\}\}$, and player~2’s partition is $\Pi_2=\{\{\omega_1\},\{\omega_2\},\{\omega_3,\omega_4\}\}$.  Thus, when either $\omega_1$ or $\omega_2$ is realized, player~2 learns the state with probability~1, while player~1’s posterior is uniform over $\{\omega_1,\omega_2\}$.  Similarly, when $\omega_3$ or $\omega_4$ is realized, then player~1 learns the state with probability~1, while player~2’s posterior is uniform over $\{\omega_3,\omega_4\}$.

The payoffs of the game are presented in Figure~\ref{Figure - negotiation games matrices}. Let $G_i$ denote the payoff matrix when state $\omega_i$ is realized.  In $G_1$ and $G_3$, player~1’s dominant strategy is $A$ and player~2’s dominant strategy is $C$, whereas in $G_2$ and $G_4$ the dominant strategies are reversed.

\begin{figure}[th!]
\centering
\setlength{\tabcolsep}{12pt}
\renewcommand{\arraystretch}{1.4}

\[
\begin{array}{c}
\begin{array}{c c| cc|}
      &      & \multicolumn{2}{c|}{\textbf{Player 2}} \\[-0.3em]
      &      & A & C \\ \cline{1-4}
\multirow{2}{*}{\textbf{Player 1}} 
      & \multicolumn{1}{c|}{A} & (2,-5)   & (2,-4) \\
      & \multicolumn{1}{c|}{C} & (-1,-1) & (0,0) \\ \cline{1-4}
\end{array} \\[0.5em]
\text{Given that } \omega \in \{\omega_1, \omega_3\}
\end{array}
\qquad\qquad
\begin{array}{c}
\begin{array}{c c| cc|}
      &      & \multicolumn{2}{c|}{\textbf{Player 2}} \\[-0.3em]
      &      & A & C \\ \cline{1-4}
\multirow{2}{*}{\textbf{Player 1}} 
      & \multicolumn{1}{c|}{A} & (-5,2)  & (-1,-1) \\
      & \multicolumn{1}{c|}{C} & (-4,2)  & (0,0) \\ \cline{1-4}
\end{array} \\[0.5em]
\text{Given that }  \omega \in \{\omega_2, \omega_4\}
\end{array}
\]

%\caption{Two payoff matrices with Player~1 (row player) and Player~2 (column player). If $\omega_1$ or $\omega_2$ are realized, then $x=2$, and if $\omega_3$ or $\omega_4$ are realized, then $x=3$.}
\caption{Two payoff matrices with Player~1 (row player) and Player~2 (column player). }
\label{Figure - negotiation games matrices}
\end{figure}

The interpretation of the game is straightforward.  Each player can either attack the other or compromise by agreeing to a peace treaty.  In states $\omega_1$ and $\omega_3$, player~1 holds the superior attacking position, whereas in states $\omega_2$ and $\omega_4$ player~2 holds this advantage. Nevertheless, from a social perspective, it might be optimal to reach a peace agreement, as it maximizes the aggregate payoff.  Such a treaty, however, requires a joint concession by both players.

Consider now the players’ equilibrium behavior. 
If the realized state is either $\omega_1$ or $\omega_2$, then player~2 is fully informed while player~1’s posterior is $\left(1/2,1/2,0,0\right)$.  In this case, given that player~$2$'s dominant strategies are $C$ in $\o_1$ and $A$ in $\o_2$, player~$1$’s optimal action is $A$.  On the other hand, if the realized state is either $\omega_3$ or $\omega_4$, the roles reverse, and player~2’s optimal action is $A$.  Hence, in every equilibrium of the game, the profile $(C,C)$ is never played.

Now consider the role of a mediator. There are two possible mediators, indexed $j=1,2$, and each possesses private information. Mediator $1$ is represented by the partition $F_1=\{\{\omega_1,\omega_3\},\{\omega_2,\omega_4\}\}$, and Mediator $2$ is represented by the partition $F_2=\{\{\omega_1,\omega_4\},\{\omega_2,\omega_3\}\}$. Figure~\ref{fig: negotiation game information structure} depicts the players’ and the mediators' information structures.
Notably, conditional on any CKC, both mediators have perfect information of the realized state. Thus, in every CKC they can induce any Bayes-plausible posterior profile. Yet, their ability to induce equilibrium outcomes differs substantially. 

\begin{figure}[th!]
\centering
\begin{minipage}{0.45\textwidth}
\centering
\begin{tikzpicture}[scale=0.8]

% Draw the outer rectangle
\draw[thick] (0,0) rectangle (6,6);

% Draw Omega label
\node at (0.3,5.7) {$\Omega$};

% Draw the Pi_1 label in blue
\node[blue] at (2,5.6) {$\Pi_1$};

% Draw the Pi_2 label in red
\node[red] at (5,5.6) {$\Pi_2$};

% Draw the large ellipse on the left in blue
\draw[blue, thick] (1.5,3) ellipse (1 and 2.5);

% Draw the small ellipse around omega_1 in red
\draw[red, thick] (1.55,4.5) ellipse (0.6 and 0.6);

% Draw the small ellipse around omega_2 in red
\draw[red, thick] (1.55,1.5) ellipse (0.6 and 0.6);

% Draw the large ellipse on the right in red
\draw[red, thick] (4.5,3) ellipse (1 and 2.5);

% Draw the small ellipse around omega_3 in blue
\draw[blue, thick] (4.55,4.5) ellipse (0.6 and 0.6);

% Draw the small ellipse around omega_4 in blue
\draw[blue, thick] (4.55,1.5) ellipse (0.6 and 0.6);

% Draw dots for omega_1, omega_2, omega_3, and omega_4
\filldraw[black] (1.35,4.5) circle (2pt) node[anchor=west] {$\o_1$};
\filldraw[black] (1.35,1.5) circle (2pt) node[anchor=west] {$\o_2$};
\filldraw[black] (4.35,4.5) circle (2pt) node[anchor=west] {$\o_3$};
\filldraw[black] (4.35,1.5) circle (2pt) node[anchor=west] {$\o_4$};

\end{tikzpicture}
\caption*{(a)}
\vspace{-0.2cm}
\caption*{The players' information}
\end{minipage}%
\hfill
\begin{minipage}{0.45\textwidth}
\centering
\begin{tikzpicture}[scale=0.8]

% Draw the outer rectangle
\draw[thick] (0,0) rectangle (6,6);

% Draw Omega label
\node at (0.3,5.7) {$\Omega$};

% Draw the F_2 label in orange
\node[orange] at (3,3) {$F_2$};

% Draw the F_1 label in teal
\node[teal] at (3,5) {$F_1$};

% Draw the large ellipse on the right in blue
\draw[teal, thick, rotate around={90:(3,3)}] (4.5,3) ellipse (0.8 and 2.5);

% Draw the diagonal ellipse around omega_1 and omega_4 in purple
\draw[teal, thick, rotate around={90:(3,3)}] (1.5,3) ellipse (0.8 and 2.5);

% Draw the diagonal ellipse around omega_1 and omega_4 in orange
\draw[orange, thick, rotate around={45:(3,3)}] (3,3) ellipse (0.8 and 3);

% Draw the diagonal ellipse around omega_1 and omega_4 in orange
\draw[orange, thick, rotate around={135:(3,3)}] (3,3) ellipse (0.8 and 3);

% Draw dots for omega_1, omega_2, omega_3, and omega_4
\filldraw[black] (1.5,4.5) circle (2pt) node[anchor=west] {$\o_1$};
\filldraw[black] (1.5,1.5) circle (2pt) node[anchor=west] {$\o_2$};
\filldraw[black] (4.3,4.5) circle (2pt) node[anchor=west] {$\o_3$};
\filldraw[black] (4.3,1.5) circle (2pt) node[anchor=west] {$\o_4$};

\end{tikzpicture}
\caption*{(b)}
\vspace{-0.2cm}
\caption*{The mediators' information}
\end{minipage}
\caption{On the left, Figure (a) illustrates the information structure of player $1$ (blue) and player $2$ (red). On the right, Figure (b) portrays the information structure of the Mediator 1 (teal) and Mediator 2 (orange).}
\label{fig: negotiation game information structure}
\end{figure}

Assume, for simplicity, that each mediator tries to persuade both players to accept a peace treaty in equilibrium. This can never succeed in states $\o_2$ and $\o_3$, since in those states the fully informed player holds the superior attacking position and necessarily chooses $A$. Thus peace can be induced only in states $\o_1$ and $\o_4$, where the fully informed player is willing to compromise, and the uninformed player must be persuaded to do the same.

The relevant incentive constraints are simple. In $\{\o_1,\o_2\}$, player~1 is willing to play $C$ iff his posterior probability of $\o_1$ is at most $1/3$. In $\{\o_3,\o_4\}$, player~2 is willing to play $C$ iff his posterior probability of $\o_3$ is at least $2/3$. Hence, a signal that supports peace must make the first uncertain player sufficiently pessimistic about having the attacking advantage, and the second uncertain player sufficiently optimistic that the other player has it.

This is where the two mediators differ. Under $F_1$, any signal that induces posterior $(p,1-p,0,0)$ for player~1 also induces posterior $(0,0,p,1-p)$ for player~2. Hence, the same signal can support peace in both regions only if $p\leq 1/3$ and $p\geq 2/3$, which is impossible. Thus Mediator~1 cannot use a single signal to make both uncertain players willing to compromise. He can still induce peace by using different signals for the two regions, but this is somewhat inefficient, as the induced probability of peace is bounded by $1/6$.\footnote{Let \(x=\mathbb{P}((C,C)\mid\omega_1)\) and \(y=\mathbb{P}((C,C)\mid\omega_4)\). By \(F_1\)-measurability and the incentive constraints, a signal supporting peace at \(\omega_1\) (or at \(\omega_4\)) must be at least twice as likely at \(\omega_2\) (at \(\omega_3\), respectively). Since no signal supports peace in both regions, \(2x+y\leq1\) and \(x+2y\leq1\). Thus the ex ante peace probability is at most \((x+y)/4\leq1/6\).}

By contrast, under $F_2$, any signal that induces posterior $(p,1-p,0,0)$ for player~1 induces posterior $(0,0,1-p,p)$ for player~2. Now the two incentive constraints coincide: the same signal supports peace in both regions whenever $p\leq 1/3$. For example, a signal sent with probability $1/2$ in states $\o_1$ and $\o_4$, and with probability $1$ in states $\o_2$ and $\o_3$, induces posteriors $(1/3,2/3,0,0)$ and $(0,0,2/3,1/3)$. After this signal, both uncertain players are willing to play $C$, so there is an equilibrium in which $(C,C)$ is played in states $\o_1$ and $\o_4$ whenever the signal is sent. This yields an ex ante probability of peace equal to $1/4$.

Consequently, the difference between the mediators does not lie in the information they possess within each CKC, as they are equally informative in that respect. The difference is how their information ``glues" the two CKCs together. Mediator~1 glues the states in a way that makes the two incentive constraints incompatible, while Mediator~2 glues them in a way that makes them coincide. Hence, two mediators with the same perfect information in every CKC can induce different equilibrium outcomes. Indeed, our results would show that the mediators are not equivalent, and that neither dominates the other. The fact-checking example in Appendix~\ref{Appendix: polarization example} gives a similar loop mechanism, by showing that even when a partition is \emph{finer inside every CKC}, it can still be inferior in mitigating differences in the posterior beliefs of players, simply because its atoms are globally glued differently.\footnote{The information structure underlying the mediation game also admits an alternative economic interpretation in the context of a takeover contest. In that setting, the two oracles can be viewed as outside analysts whose public disclosures affect whether competing bidders engage in a bidding war. Appendix \ref{Appendix: takeover bidding war} develops this interpretation and shows that two analysts who are equally informative within each CKC may nevertheless support different bidding outcomes because their information connects the CKCs differently.}

Finally, consider a third mediator with the partition $F_3=\{\{\o_1\},\{\o_4\},\{\o_2,\o_3\}\}$. This mediator reveals the states in which peace can be induced and pools the states in which peace is impossible. It can use the same experiment as Mediator $2$ above. The induced probability of peace remains $1/4$, which is the highest possible since in each of the states $\o_1$ and $\o_4$, the relevant likelihood of the peace-inducing signal is at most $1/2$.

This third mediator is useful because it separates the game-theoretic point from the informational one. While $F_3$ is not itself the full-information partition, our characterization (Theorem~\ref{Theorem: Equivalent oracles}) implies that it is equivalent to a perfectly informed oracle: it imposes no loop restriction beyond Bayes plausibility. Theorem~\ref{Theorem: stochastic loop cover dominance} and Proposition~\ref{Proposition: two CKCs} below imply that $F_3$ dominates not only $F_2$, which it refines, but also $F_1$, despite not refining it.

%--------------------------------------------------------------------------
%--------------------------------------------------------------------------
\section{The model and posterior mimicry} \label{Section - Model}
%--------------------------------------------------------------------------
%--------------------------------------------------------------------------

%****** Outline in comments ******

A \emph{guided game} comprises a Bayesian game and an \emph{oracle}. The oracle's role is to provide information that enables a different, and preferably broader, range of equilibria. It does so through public signaling, and our analysis seeks to characterize the extent to which oracles can expand the set of equilibrium payoffs.

%****** The Bayesian game ******

We first define the underlying Bayesian game. Let \( N = \{1, 2, \dots, n\} \) be a finite set of \( n \geq 2 \) players, and let \( \Omega \) denote a non-empty, finite state space. Each player \( i \in N \) has a non-empty finite set of actions \( A_i \) and a partition \( \Pi_i \) over \( \Omega \), representing the information available to player \( i \). Denote the set of action profiles by \( A = \times_{i \in N} A_i \). The utility function for each player \( i \in N \) is \( u_i: \Omega \times A \to \mathbb{R} \), which maps states and action profiles to real-valued payoffs.\footnote{The underlying state space $\Omega$ represents payoff-relevant fundamentals only, referred to in \cite{Maschler2013} as the states of nature. Agents’ information is modeled by their information partitions, and payoffs do not depend directly on agents’ information or beliefs. The states of nature together with these partitions and prior, when they are common knowledge among all players, determine the players’ types, which are characterized by the full hierarchy of beliefs, à la Harsanyi.}

The players' partitions determine a common-knowledge partition. Following \cite{Aumann1976}, a \emph{common knowledge component} (CKC) is an atom of the meet \(\bigwedge_{i=1}^n \Pi_i\), the finest common coarsening of the players' partitions. Equivalently, a CKC is a minimal event, with respect to set inclusion, that is common knowledge among the players. We denote the set of CKCs by \(\mathcal C\), and for each \(\omega\in\Omega\), let \(C(\omega)\in\mathcal C\) be the CKC containing \(\omega\). 

%****** The Oracle and its strategy ******

To extend the basic game into a guided game, we introduce an oracle that provides public information before players choose their actions. The oracle is endowed with a partition $F$ of the state space $\Omega$, and a countable set $S$ of possible signals. A \textit{signaling function} of the oracle is an $F$-measurable experiment $\tau : \Omega \to \Delta(S)$, where $\Delta(S)$ is the set of probability distributions on $S$ with finite support. We denote by $\tau(s|\omega)$ the probability that the oracle sends signal $s$ when the realized state is $\omega$.

%****** The evolution of the game ******

The guided game evolves as follows. First, the oracle publicly announces an experiment $\tau$. Then, a state $\omega \in \Omega$ is drawn according to a common prior $\mu \in \Delta (\Omega)$ with full support. Each player $i$ is privately informed of $\Pi_i(\omega)$, which is a set of states containing $\omega$ and also an atom of player $i$'s private partition. Finally, a signal $s\in S$ is drawn according to $\tau(\omega)$ and is publicly announced.

Let $\mu^i_{\tau\mid\omega,s}= \mu(\cdot\mid\Pi_i(\omega),\tau,s) \in \Delta(\Omega)$ denote player $i$'s posterior belief after observing $\Pi_i(\omega)$ and
a realized signal $s$ according to $\tau(\omega)$, whenever this observation has positive probability. The experiment \(\tau\), together with the prior and the players' partitions, induces a finite set \(\operatorname{Post}(\tau)\) of realized joint posterior profiles. The prior, the players' partitions, and the experiment \(\tau\) induce the guided incomplete-information game \(G(\tau)\). When there is no risk of ambiguity, we denote the incomplete-information game without \(\tau\) by \(G\).

\subsection{Comparing Oracles: Dominance and Equivalence Classes}

To discuss the role of the oracle in the current framework, we define a relevant solution concept, referred to as a \emph{Guided equilibrium}, which incorporates the oracle's strategy.
Formally, let $\sigma_i: \Pi_i\times S\rightarrow \Delta (A_i)$ be a strategy of player $i$. A tuple $(\tau,\sigma_1,\dots,\sigma_n)$ is a \emph{Guided equilibrium} if $(\sigma_1,\dots,\sigma_n)$ is a Nash equilibrium in the incomplete-information game $G(\tau)$.

The notion of a Guided equilibrium defines a \emph{preorder} of oracles, i.e., a reflexive and transitive relation over their partitions according to the sets of equilibria. To define this relation, let $\rm{NED}(G(\tau)) \subseteq \Delta(\Omega \times A)$ be the set of distributions over $\Omega \times A$ induced by Nash equilibria given $G$ and $\tau$.\footnote{Note that a Nash equilibrium $(\sigma_i^*,...,\sigma_n^*)$ induces a probability distribution over $\Omega \times A$. Specifically, fix $\omega$ and an action profile $a$, the probability of $(\omega,a)$ under the equilibrium strategy $(\sigma_i^*,...,\sigma_n^*)$ and the experiment $\tau$ is given by $\mu(\omega)\sum_{s\in S} \tau(s|\omega)\prod_{i=1}^{n} \sigma^*_i(a_i|\Pi_i(\omega),s)$. Since multiple equilibria can exist, $\rm{NED}(G(\tau))$ is a subset of $\Delta(\Omega \times A)$.}
Now consider two oracles, Oracle $1$ and Oracle $2$, and denote the generic partition and experiment of Oracle $j$ by $F_j$ and $\tau_j$, respectively.

\begin{definition}[Dominance and equivalence of oracles] \label{Definition - Strategic Dominance}
    Fix the players' information structures. We say that \emph{Oracle $1$ dominates Oracle $2$}, denoted $F_1 \succeq_{\rm{NE}} F_2$, if for every $\tau_2$, there exists $\tau_1$ such that for every game $G$, we have $\rm{NED}(G(\tau_2)) \subseteq \rm{NED}(G(\tau_1))$. We say that \emph{Oracle $1$} and \emph{Oracle $2$} are \emph{equivalent}, denoted $F_1 \sim_{\rm{NE}} F_2$, if $F_i \succeq_{\rm{NE}} F_{-i}$, for every $i=1,2$.
\end{definition}

Oracle 1 dominates Oracle 2 if, for each experiment chosen by Oracle 2, Oracle 1 can choose a single experiment that implements, in every game, all equilibrium outcome distributions implementable under Oracle 2's experiment. A direct comparison of players' equilibrium strategies is ill-defined across oracles, as their domains are endogenously determined by the chosen experiment.

Three points are worth noting here. First, one could consider defining dominance between oracles in a more robust manner by allowing players' information structures to vary over a set of possible partitions. If dominance were required to hold uniformly over all such partitions, the comparison problem would become rather simple. In particular, the set of admissible information structures would include the case of trivial private information. In that case, the common-knowledge structure admits no loops and, by our results, oracle dominance reduces to partition refinement. The substantive challenge in our framework arises precisely because players' information partitions are fixed and predetermined, so that dominance must be assessed under binding measurability and common-knowledge constraints.

Second, Definition \ref{Definition - Strategic Dominance} compares oracles via inclusion of equilibrium outcome distributions. Analogously, one can define dominance in terms of equilibrium payoff sets. Outcome–based dominance implies the corresponding payoff–based notion, but the former is more natural when the objective is to control actions and aggregate outcomes, whereas payoff–based notions suit environments that focus on agents’ utilities. Evidently, Oracle~1 may support additional equilibrium outcomes.

Third, one can compare oracles by treating them as strategic players in some sender--receiver game, assigning payoff functions, and saying that Oracle~1 is more informative than Oracle~2 if it obtains weakly higher equilibrium payoffs in every such game. This approach faces two difficulties: (i) equilibria are typically multiple, so oracle payoffs depend on an arbitrary selection rule; and (ii) it ties the comparison of information structures to particular oracle objectives. Our dominance notion is instead an analyst's environment-specific implementability comparison. The players' information partitions are fixed, and the matching experiment may depend on that configuration; an operational rule for selecting it would therefore generally require knowledge of the fixed information structure. The formal relation itself, however, neither assigns an objective to the oracle nor requires the oracle to know those partitions.

\subsection{A CKC-posterior characterization of dominance} \label{Section:characterization using oracles' strategies}

The following proposition reduces strategic dominance to posterior mimicry within CKCs. For an experiment \(\tau\), a CKC \(C\), and a signal \(s\) with \(\mathbb P_\tau(s\mid C) := \sum_{\omega\in C}\mu(\omega\mid C)\tau(s\mid\omega)>0,\) let \(P_{\tau,s,C}\in\Delta(C)\) denote the posterior on \(C\) induced by \(s\), and let \(Q_\tau(C):=\{P_{\tau,s,C}:\mathbb P_\tau(s\mid C)>0\}\). Let \(\overline{\mu}_{\tau\mid C}\) denote the distribution of \(P_{\tau,s,C}\) when \(s\) is drawn according to \(\mathbb P_\tau(\cdot\mid C)\). Since players' information partitions are fixed, the CKC-posterior \(P_{\tau,s,C}\) also determines each player's posterior within \(C\) by conditioning on her information cell. (All main proofs are deferred to Appendix~\ref{Appendix: proofs}.)

\begin{proposition} \label{Prop - Charact in terms of signaling funct}
Oracle~$1$ dominates Oracle~$2$ if and only if for every experiment $\tau_2$ there exists an experiment $\tau_1$ such that $\overline{\mu}_{\tau_1 \mid C}= \overline{\mu}_{\tau_2 \mid C}$ for every CKC $C$.
\end{proposition}

The equality \(\overline{\mu}_{\tau_1 \mid C}= \overline{\mu}_{\tau_2 \mid C}\) matches the distribution of posterior beliefs within each CKC, but it need not preserve the signal-label structure of \(\tau_2\). This matters because payoff-irrelevant public labels can serve as coordination devices in equilibrium. In the proof, Oracle~1 restores this coordination structure by augmenting its experiment with state-independent common randomness that selects, conditional on each posterior \(q\in Q_{\tau_2}(C)\), among the \(\tau_2\)-signals inducing \(q\) with the same conditional probabilities as under \(\tau_2\). In particular, Oracle~1 dominates Oracle~2 if every $F_2$-measurable experiment can be reproduced inside every CKC, up to common noise, by a single $F_1$-measurable experiment.

\section{Information loops} \label{Section: Information loops}

Our characterization of dominance has two components: a local one, comparing the oracles' partitions within each CKC, and a global one, capturing how the two partitions link different CKCs. To analyze the global component, we develop a theory of \emph{information loops}, presented below.

\subsection{Local refinement inside CKCs}

The local component of dominance compares the two oracles within each CKC. For a partition $F$ and a CKC $C$, write $F|_C$ for the restriction of $F$ to $C$, namely the collection of non-empty sets $A\cap C$ with $A\in F$. We say that $F_1$ \emph{refines} $F_2$ in the CKC $C$ if $F_1|_C$ refines $F_2|_C$, and we say that $F_1$ refines $F_2$ in every CKC if this condition holds for every CKC. This local refinement condition is the within-component part of dominance. The remaining difficulty is global: a single $F_1$-measurable experiment must be chosen across all CKCs, and this creates additional measurability constraints.

When there is a single CKC, dominance is essentially local: refinement inside that component is enough to reproduce the relevant posterior distributions. With several CKCs, this local argument no longer settles the problem. Atoms of the oracle partition may connect different CKCs, and the mimicking experiment must be measurable across all these links simultaneously. The compatibility constraints generated by such cross-component links are the source of the loop restrictions below.
Appendix~\ref{Subsection_more than one CKC two examples} gives examples showing that, with several CKCs, a mimicking oracle may need to use a different signal space than the oracle it mimics, even when the two oracles have the same information inside each CKC.

\subsection{Information loops and covers} \label{Section - Information Loops}

We now define the loop objects that create global constraints across CKCs.
Let \(C_1,\dots,C_l\) be the CKCs. A key aspect of our analysis is that atoms of an oracle partition may connect different CKCs. A sequence of such connections can form a closed path, which we call an information loop, or simply a loop. 

\begin{definition} \label{Definition: a loop}
    An $F_i$-loop is a sequence $(\omega_1,\overline{\omega}_1, \omega_2,\overline{\omega}_2, \dots, \omega_m,\overline{\omega}_m)$ of $2m$ distinct states, with $m\geq 2$ where indices are taken modulo $m$, such that
    \begin{itemize}
        \item [a.]   $\omega_j,\overline{\omega}_j \in C_{r_j}$ for all $j=1,\dots,m$;\footnote{Here $C_{r_j}$ refers to the CKC that contains the $j$-th pair of states $(\omega_j,\overline{\omega}_j)$.}
        \item [b.] $\omega_{j+1} \in F_i(\overline{\omega}_j)$ for all $j=1,\dots,m$;
        \item [c.] $C_{r_j} \neq C_{r_{j+1}}$ for all $j=1,\dots,m$.
    \end{itemize}
\end{definition}

%%%%%%%%%%%%%%%%%%%%%%%%%%%%%

The concept of an information loop captures a strategic tension between the players' common-knowledge structure and the oracle's information partition. An information loop links distinct CKCs because the oracle need not know what is common knowledge among the players. Consequently, its partition cells may cross CKC boundaries. Formally, an $F_i$-loop relies on the property that two states belonging to different CKCs, namely $\overline{\omega}_j$ and $\omega_{j+1}$, satisfy
$
F_i(\overline{\omega}_j)=F_i(\omega_{j+1}).
$
Since the oracle cannot distinguish between these states, any $F_i$-measurable signal structure must treat them identically, generating global consistency restrictions across distinct CKCs.

At the same time, a loop may contain pairs of states that lie within the same CKC, namely, that the players cannot commonly distinguish, while the oracle can. Accordingly, each pair $(\omega_j,\overline{\omega}_j)$ in the loop consists of states that may belong to different information sets of $F_i$, even though they are drawn from the same CKC. Information loops, therefore, arise from the following tension:
(a) The oracle identifies states across different CKCs, because it does not observe the players' common-knowledge boundaries;
    (b) The oracle may distinguish states within a CKC that are not separated by common knowledge among the players.

Information loops admit a useful graph-theoretic interpretation. Consider the CKCs as the vertices of a multigraph, and connect two CKCs whenever there exist states $\overline{\omega}_j$ and $\omega_{j+1}$ in the same $F_i$-partition element. An information loop then corresponds to a closed walk through selected cross-CKC links. The condition $C_{r_j}\neq C_{r_{j+1}}$ ensures that attention is restricted to genuinely cross-CKC links, since these are precisely the links generating global compatibility constraints. %Examples appear in Figures~\ref{fig: F1 irreducible loop} and~\ref{Figure: covered by F2 loops}.

The importance of loops follows directly from measurability. If $\tau_i$ is $F_i$-measurable and $(\omega_1,\overline{\omega}_1,\dots,\omega_m,\overline{\omega}_m)$ is an $F_i$-loop, then for every fully supported signal $s$,
\begin{equation}\label{eq: loop-product-constraint}
    \prod_{j=1}^m \frac{\tau_i(s\mid \omega_j)}{\tau_i(s\mid \overline{\omega}_j)}=1.
\end{equation}
Indeed, since $\overline{\omega}_j$ and $\omega_{j+1}$ belong to the same $F_i$-atom, the denominator in one term equals the numerator in the next. Thus, every $F_i$-measurable experiment must satisfy the loop-product restriction induced by the corresponding information loops.

The central question is therefore whether every $F_2$-measurable experiment that can be locally replicated within each CKC also satisfies the loop restrictions imposed by $F_1$. We address this question through the notion of \emph{covers}.

\begin{definition} \label{Definition - covered loop}
    An $F_i$-loop $(\omega_1,\overline{\omega}_1, \omega_2,\overline{\omega}_2, \dots, \omega_m,\overline{\omega}_m)$ is $F_{-i}$\emph{-covered} if the set $\{1,\dots,m\}$ can be partitioned into disjoint sets of indices $J,I_1,\dots,I_r$ such that $\big((\omega_j,\overline{\omega}_j) \big)_{j\in I_t}$ yields an $F_{-i}$-loop for each $t=1,\dots,r$ (also referred to as an $F_{-i}$-\emph{sub-loop}) and $\omega_j\in F_{-i}(\overline{\omega}_j)$ for every $j\in J$.\footnote{The order of the pairs $(\omega_j,\overline{\omega}_j)$ in the $F_{-i}$-loop need not coincide with their order in the original $F_i$-loop. For instance, an $F_1$-loop $(\omega_1,\overline{\omega}_1,\omega_2,\overline{\omega}_2,\omega_3,\overline{\omega}_3)$ might be covered by the $F_2$-loop $(\omega_1,\overline{\omega}_1,\omega_3,\overline{\omega}_3,\omega_2,\overline{\omega}_2)$. Whenever $F_{-i}(\omega_j) = F_{-i}(\overline{\omega}_j)$, Oracle $-i$ cannot distinguish between $\omega_j$ and $\overline{\omega}_j$, so the pair may be placed in $J$. \label{Footnote - order of subloop}}
    The cover is \emph{order-preserving} if every $F_{-i}$-sub-loop in the cover follows the same cyclic ordering of pairs as the original $F_i$-loop, up to a cyclic rotation. An $F_{-i}$-non-informative pair need not belong to $J$; it may instead be retained in an $F_{-i}$-sub-loop.
\end{definition}

In intuitive terms, an \(F_{-i}\)-cover decomposes an \(F_i\)-loop into a collection of \(F_{-i}\)-loops together with pairs for which Oracle~\(-i\) is non-informative. Figure~\ref{Figure: covered by F2 loops} illustrates these definitions. In all panels, each rectangle represents a CKC, and the underlying structure is the same \(F_1\)-loop, namely \(((\omega_j,\overline{\omega}_j))_{j=1,\dots,4}\). The panels differ only in the informational structure induced by \(F_2\). Panel~(a) depicts a cover consisting of two order-preserving \(F_2\)-loops. Panel~(b) presents a partition \(F_2\) that fails to cover the original \(F_1\)-loop: \(\overline{\omega}_1\) is matched with \(\overline{\omega}_3\) rather than with \(\omega_3\). The panel also contains two \(F_2\)-non-informative pairs, namely \((\omega_2,\overline{\omega}_2)\) and \((\omega_4,\overline{\omega}_4)\). Panel~(c) illustrates a cover that is not order-preserving. The sequence
\(
(\omega_1,\overline{\omega}_1,\omega_3,\overline{\omega}_3,\omega_2,\overline{\omega}_2)
\)
forms an \(F_2\)-loop, while \((\omega_4,\overline{\omega}_4)\) is an \(F_2\)-non-informative pair. In this case, the pair \((\omega_3,\overline{\omega}_3)\) appears before \((\omega_2,\overline{\omega}_2)\), thereby violating the original ordering induced by the \(F_1\)-loop. Our main theorems demonstrate that the order-preservation requirement is essential.

\begin{figure}[ht!]
\centering

\begin{tikzpicture}[scale=0.8]

    % Replace blue lines with ellipses
    \draw[blue, thick] (2.2, -1.25) ellipse (0.5 and 3);  % First blue ellipse
    \draw[teal, thick] (3.8, -1.25) ellipse (0.5 and 3);  % Second blue ellipse

    % Replace red lines with ellipses
    \draw[red, thick] (3, -0.5) ellipse (2.5 and 0.5);  % First red ellipse
    \draw[orange, thick] (3, -2) ellipse (2.5 and 0.5);    % Second red ellipse

    % Draw the rectangles with labels
    \draw (1.5,0.5) -- (4.5,0.5) -- (4.5,1.5) -- (1.5,1.5) -- cycle;
    \node at (2.2,1) {$\o_1$};

    \node at (3.8,1) {$\overline{\o}_1$};

    \draw (4.5,0) -- (5.5,0) -- (5.5,-2.5) -- (4.5,-2.5) -- cycle;
    \node at (5,-0.5) {$\o_2$};

    \node at (5,-2) {$\overline{\o}_2$};

    \draw (0.5,0) -- (1.5,0) -- (1.5,-2.5) -- (0.5,-2.5) -- cycle;
    \node at (1,-2) {$\o_4$};

    \node at (1,-0.5) {$\overline{\o}_4$};

    \draw (1.5,-3.0) -- (4.5,-3.0) -- (4.5,-4.0) -- (1.5,-4.0) -- cycle;
    \node at (2.2,-3.5) {$\overline \o_3$};

    \node at (3.8,-3.5) {${\o_3}$};

    \node[blue] at (2.2,2.1) {$F_2(\o_1)$};
    \node[teal] at (3.8,2.1) {$F_2(\o_3)$};
    \node[red] at (6.3,-0.5) {$F_2(\o_2)$};
    \node[orange] at (6.3,-2) {$F_2(\o_4)$};
%%%%%%%%%%%%%%%%%%%%%%%%%%%%%%%%%%%%%%%%%%%%%%%%%%%%%%%%%%%%%%%%%%%%%%%%%
    % Replace blue lines with ellipses
%    \draw[blue, thick] (11.2, -1.25) ellipse (0.5 and 3);  % First blue ellipse
%    \draw[teal, thick] (12.8,-1.25) ellipse (0.5 and 3);  % Second blue ellipse

    \draw[blue, thick, rotate around={-69:(10,-1.25)}] (10,-1.25) ellipse (3.0cm and 0.7cm);
    \draw[teal, thick, rotate around={69:(10,-1.25)}] (10,-1.25) ellipse (3.0cm and 0.7cm);

    % Replace red lines with ellipses
    \draw[red, thick] (8, -1.25) ellipse (0.6 and 2);  % First red ellipse
    \draw[orange, thick] (12, -1.25) ellipse (0.6 and 2);    % Second red ellipse

    % Draw the rectangles with labels
    \draw (8.5,0.5) -- (11.5,0.5) -- (11.5,1.5) -- (8.5,1.5) -- cycle;
    \node at (9.2,1) {$\o_1$};

    \node at (10.8,1) {$\overline{\o}_1$};

    \draw (11.5,0) -- (12.5,0) -- (12.5,-2.5) -- (11.5,-2.5) -- cycle;
    \node at (12,-0.5) {$\o_2$};

    \node at (12,-2) {$\overline{\o}_2$};

    \draw (7.5,0) -- (8.5,0) -- (8.5,-2.5) -- (7.5,-2.5) -- cycle;
    \node at (8,-2) {$\o_4$};

    \node at (8,-0.5) {$\overline{\o}_4$};

    \draw (8.5,-3.0) -- (11.5,-3.0) -- (11.5,-4.0) -- (8.5,-4.0) -- cycle;
    \node at (9.2,-3.5) {$\overline \o_3$};

    \node at (10.8,-3.5) {${\o_3}$};

    \node[blue] at (9,2.1) {$F_2(\o_1)$};
    \node[teal] at (10.8,2.1) {$F_2(\overline{\o}_3)$};
    \node[orange] at (12.3,1.2) {$F_2(\o_2)$};
    \node[red] at (7.7,1.2) {$F_2(\o_4)$};

%%%%%%%%%%%%%%%%%%%%%%%%%%%%%%%%%%%%%%%%%%%%%%%%%%%%%%%%%%%%%%%%%%%%%%%%%
    % Replace blue lines with ellipses
    \draw[blue, thick] (17.8, -1.25) ellipse (0.5 and 3);  % First blue ellipse

    \draw[teal, thick, rotate around={-45:(17.6,-0.55)}] (17.6,-0.55) ellipse (3.0cm and 0.6cm);
    \draw[orange, thick, rotate around={45:(17.6,-2)}] (17.6,-2) ellipse (3.0cm and 0.6cm);

    % Replace red lines with ellipses
    \draw[red, thick] (15, -1.25) ellipse (0.6 and 2);  % First red ellipse

    % Draw the rectangles with labels
    \draw (15.5,0.5) -- (18.5,0.5) -- (18.5,1.5) -- (15.5,1.5) -- cycle;
    \node at (16.2,1) {$\o_1$};

    \node at (17.8,1) {$\overline{\o}_1$};

    \draw (18.5,0) -- (19.5,0) -- (19.5,-2.5) -- (18.5,-2.5) -- cycle;
    \node at (19,-0.5) {$\o_2$};

    \node at (19,-2) {$\overline{\o}_2$};

    \draw (14.5,0) -- (15.5,0) -- (15.5,-2.5) -- (14.5,-2.5) -- cycle;
    \node at (15,-2) {$\o_4$};

    \node at (15,-0.5) {$\overline{\o}_4$};

    \draw (15.5,-3.0) -- (18.5,-3.0) -- (18.5,-4.0) -- (15.5,-4.0) -- cycle;
    \node at (16.2,-3.5) {$\overline \o_3$};

    \node at (17.8,-3.5) {${\o_3}$};

    \node[teal] at (15.7,2) {$F_2(\o_1)$};
    \node[blue] at (17.8,2) {$F_2(\overline{\o}_1)$};
    \node[orange] at (19.6,0.6) {$F_2(\o_2)$};
    \node[red] at (14.7,1.2) {$F_2(\o_4)$};

    \node at (3,-5) {(a)};
    \node at (10,-5) {(b)};
    \node at (17,-5) {(c)};

\end{tikzpicture}

\caption{\footnotesize All panels show an $F_1$-loop, consisting of $((\o_j,\overline{\o}_j))_{j=1,\dots,4}$, with different \(F_2\)-configurations relative to the same
\(F_1\)-loop.} \label{Figure: covered by F2 loops}
\end{figure}

\subsection{Irreducible and informative loops}

The ability to classify covers, and specifically order-preserving covers, is rather difficult. It appears that there is currently no relevant theory for this classification. To face these challenges, we decompose loops and covers into smaller objects, referred to as \emph{irreducible loops}, and use these objects as building blocks in our main results. Formally, an $F_i$-loop is irreducible if it does not contain a smaller $F_i$-loop using only states from the original loop.

\begin{definition} \label{Definition - irreducible loop}
    Let $L_i=(\omega_1,\overline{\omega}_1,\omega_2,\overline{\omega}_2,\dots,\omega_m,\overline{\omega}_m)$ be an $F_i$-loop.
    We say that $L_i$ is \emph{irreducible} if there is no strict subset of $\{\omega_j,\overline{\omega}_j:j=1,\dots,m\}$ that forms an $F_i$-loop.
    A cover is \emph{irreducible} if every loop in the cover is irreducible.
\end{definition}

We also distinguish loops according to whether another oracle separates the two states that lie in the same CKC-pair. The non-informative case will be used in Corollary~\ref{Theorem: NI leads to dominance} as a sufficient condition for dominance.

\begin{definition} \label{Definition: informative loops}

     An $F_i$-loop $(\omega_1,\overline{\omega}_1, \omega_2,\overline{\omega}_2, \dots, \omega_m,\overline{\omega}_m)$ is $F_k$-\emph{non-informative} if $F_k(\omega_j)=F_k(\overline{\omega}_j)$ for every $j$.
     The loop is $F_k$-\emph{fully informative} if $F_k(\omega_j)\neq F_k(\overline{\omega}_j)$ for every $j$, and it is $F_k$-\emph{informative} if this inequality holds for some $j$.
\end{definition}

If an $F_1$-loop is $F_2$-non-informative, then the loop does not pose a mimicking challenge for Oracle~1, as Oracle~2 cannot provide any relevant information in any of the pairs. Algebraically, every well-defined $F_2$-measurable likelihood ratio $\tfrac{\tau_2(s\mid\omega_j)}{\tau_2(s\mid\overline{\omega}_j)}$ equals one. Thus, the $F_1$-loop product restriction in Eq.~\eqref{eq: loop-product-constraint} is automatically satisfied. 

\section{Equivalent oracles} \label{Section - Equivalent oracles}

We begin the substantive comparison of oracles with the two-sided problem, characterizing necessary and sufficient conditions for each oracle to dominate the other. The loop geometry developed above is sharpest in this setting. Equivalence requires that the two partitions agree within every CKC and satisfy the primitive global loop constraints jointly generated by the two partitions. In particular, every irreducible $F_i$-loop must have an order-preserving $F_{-i}$-cover. Theorem~\ref{Theorem: Equivalent oracles} provides this characterization.

\begin{theorem}\label{Theorem: Equivalent oracles}
$F_1 \sim_{\rm{NE}} F_2$ if and only if, for each $i=1,2$, the partition $F_i$ refines $F_{-i}$ in every CKC and every irreducible $F_i$-loop has an order-preserving $F_{-i}$-cover.
\end{theorem}

Theorem~\ref{Theorem: Equivalent oracles} separates equivalence into a local condition and a global one. The local condition is mutual refinement within each CKC, or equivalently \(F_1|_C=F_2|_C\) for every CKC \(C\). This requirement is unavoidable: if, inside some CKC, Oracle 1 can distinguish two states that Oracle 2 merges, then the former can generate posterior odds within that CKC that the latter cannot reproduce. By designing a local game embedded within this CKC based on a strictly proper scoring rule, Oracle 1 can force an equilibrium distribution reflecting these distinct posteriors. Oracle 2, bound by its internal measurability constraint, would be forced to assign equal likelihoods to both states, rendering it incapable of mimicking this belief profile.

The global condition concerns how the oracle's atoms link different CKCs. Every \(F_i\)-loop imposes a likelihood-ratio constraint, but irreducible loops are the primitive cyclic restrictions. The theorem therefore requires an order-preserving cover only for irreducible loops. This restriction is essential: reducing a non-irreducible loop may require switching partners within a repeated CKC, so the original CKC-pairs need not themselves admit a cover in the sense of Definition~\ref{Definition - covered loop}.

To conclude, Theorem \ref{Theorem: Equivalent oracles} shows that equivalent oracles are not merely those with identical information loops, but those whose public measurability constraints map out the same geometric obstructions across the players' belief hierarchies. When these cyclic constraints align symmetrically on both sides, each experiment of either oracle admits a globally measurable mimicking experiment for the other.

At first glance, Theorem~\ref{Theorem: Equivalent oracles} may suggest that equivalence between oracles coincides with equality of informational content, in the sense that each oracle refines the other, but this is false. Consider, for example, a state space \(\Omega = \{\omega_1, \omega_2, \omega_3, \omega_4\}\), with two players whose private-information partitions are \( \Pi_1=\big\{\{\omega_1,\omega_2\},\{\omega_3\},\{\omega_4\}\big\}\), and \(\Pi_2=\big\{\{\omega_1\},\{\omega_2\},\{\omega_3,\omega_4\}\big\}\), and the oracles' partitions are \(F_1 = \big\{\{\omega_1,\omega_3\}, \{\omega_2\}, \{\omega_4\}\big\}\), and \(F_2 = \big\{\{\omega_1\}, \{\omega_3\}, \{\omega_2,\omega_4\}\big\}\). This environment admits two CKCs, namely \( \{\omega_1,\omega_2\} \) and \( \{\omega_3,\omega_4\}\). Within each CKC, both oracles induce the discrete partition and therefore fully resolve uncertainty inside the component. Nevertheless, neither \(F_1\) refines \(F_2\) nor \(F_2\) refines \(F_1\).

Despite this, Theorem~\ref{Theorem: Equivalent oracles} implies that the oracles are equivalent.
The reason is that the environment contains no information loops connecting the two CKCs. Consequently, the different cross-component identifications induced by the two oracles are strategically irrelevant. If we refine \(F_1\) further and define \(F'_1 = \big\{\{\omega_1\},\{\omega_2\},\{\omega_3\}, \{\omega_4\}\big\}\), then \(F'_1\) does refine \(F_2\). However, Theorem~\ref{Theorem: Equivalent oracles} still implies that \(F'_1 \sim_{\mathrm{NE}} F_2\). In fact, the theorem introduces a notion of structural equivalence, distinct from mutual refinement, that incorporates the players’ underlying information structure via the CKCs.

\section{One-sided dominance} \label{Section - necessary and sufficient conditions for dominance}

We can now state the main one-sided dominance result. The theorem has two parts: the first gives necessary conditions for dominance in general, and the second gives a sufficient condition when the $F_1$-loops are \emph{separated}. Formally, in the bipartite incidence graph between CKCs and $F_1$-atoms, separation means that simple cycles are vertex-disjoint; equivalently, no CKC or $F_1$-atom lies on two distinct simple cycles. Together, the two parts isolate the two forces underlying dominance: local refinement within CKCs and global compatibility of loop constraints.

\begin{theorem} \label{Theorem: stochastic loop cover dominance}
Fix partitions $F_1$ and $F_2$.
\begin{enumerate}[(a)]
    \item \emph{Necessity.} If Oracle~$1$ dominates Oracle~$2$, then:
    \begin{enumerate}[(i)]
        \item $F_1$ refines $F_2$ in every \emph{CKC};
        \item every irreducible $F_1$-loop has an $F_2$-cover; and
        \item if an irreducible $F_1$-loop has a unique $F_2$-cover, then the cover is order-preserving.
    \end{enumerate}
    \item \emph{Sufficiency.} If $F_1$ refines $F_2$ in every \emph{CKC}, every irreducible $F_1$-loop has an order-preserving $F_2$-cover, and the $F_1$-loops are separated, then Oracle~$1$ dominates Oracle~$2$.
\end{enumerate}
\end{theorem}

The necessity part shows that a dominant oracle must satisfy both local and global constraints. The local constraint is that \(F_1\) refines \(F_2\) within every CKC. The global constraint is that every irreducible \(F_1\)-loop is \(F_2\)-covered, where a cover may combine \(F_2\)-sub-loops with \(F_2\)-non-informative pairs. Proposition~\ref{proposition: balanced} shows that coverability is exactly what guarantees the scalar loop-product identity for every positive \(F_2\)-measurable likelihood vector. When the cover is unique, Lemma~\ref{Lemma - unique cover order preserving} imposes the additional requirement that it preserve cyclic order. This requirement concerns the stronger experiment-level problem of assembling signal-specific likelihood vectors into normalized probability laws that reproduce the required CKC posterior distributions, not the scalar loop-product identity itself. In the unique-cover case, the associated \(F_2\)-atom multigraph has a unique directed-cycle partition and is a bridgeless cactus, so distinct covering cycles intersect in at most one \(F_2\)-atom.

The sufficiency part shows that local refinement together with order-preserving covers of irreducible loops implies dominance whenever the \(F_1\)-loops are separated. Order-preserving covers permit the normalized single-loop laws to be constructed consistently, while the separation condition allows these local constructions to be glued across loops.

\begin{remark}
Theorem~\ref{Theorem: stochastic loop cover dominance} leaves two distinct questions open. First, when an irreducible \(F_1\)-loop has several \(F_2\)-covers, we do not know whether dominance forces at least one cover to preserve order. Second, even when every irreducible loop has an order-preserving cover, overlapping incidence cycles may prevent the corresponding local constructions from being chosen consistently. These are distinct combinatorial obstructions: the first concerns order preservation within individual covers, whereas the second concerns their global compatibility. We view the identification of these obstructions as a substantive feature of the problem: oracle dominance depends on the compatibility of cyclic likelihood-ratio constraints, both within individual covers and across overlapping incidence cycles, and resolving these issues appears to require new graph-theoretic tools.
\end{remark}

\section{Benchmark cases} \label{Section - Partial ordering of oracles}

We conclude the body of the paper with several benchmark environments in which the general constraints simplify. In unique-CKC and no $F_1$-loop structures, the global ingredient is vacuous, and dominance reduces to local refinement. Under a two-CKC benchmark, global loop geometry reduces to the existence of a cover. 

\begin{corollary}[Unique-CKC and acyclic benchmarks] \label{Corollary: stochastic benchmark cases}
Assume that either $\Omega$ comprises a unique CKC, or there is no $F_1$-loop. Then, Oracle~$1$ dominates Oracle~$2$ if and only if $F_1$ refines $F_2$ in every \emph{CKC}.
\end{corollary}

This follows from Theorem~\ref{Theorem: stochastic loop cover dominance}: a unique CKC admits no $F_1$-loop, and without such loops the cover and separation conditions are vacuous (the proof is thus omitted). Next, we record two useful special cases. The first is the case of non-informative loops, in which the loop constraints imposed by $F_1$ are already respected by $F_2$. The second is the two-CKC case, where the refinement and cover conditions are both necessary and sufficient.

\begin{corollary}[Non-informative loops] \label{Theorem: NI leads to dominance}
If $F_1$ refines $F_2$ in every \emph{CKC} and every $F_1$-loop is $F_2$-non-informative, then Oracle~$1$ dominates Oracle~$2$.
\end{corollary}

\begin{proposition} \label{Proposition: two CKCs}
Assume there are only two \emph{CKCs}. Then, Oracle~$1$ dominates Oracle~$2$ if and only if $F_1$ refines $F_2$ in every \emph{CKC} and any $F_1$-loop is $F_2$-covered.
\end{proposition}

The two-CKC case is especially tractable. Every loop can be decomposed to irreducible loops, and every irreducible loop consists of two CKC-pairs so that a cover leaves only two possibilities: either the loop is $F_2$-non-informative, or it is itself an $F_2$-loop. In both cases, the order-preservation requirement is vacuous.

\section{Conclusion}\label{Section - Conclusion}
This paper develops a comparative theory of generators of public information in incomplete-information games. Fixing the players' information structures and the prior, we compare oracles by their ability to support one another's equilibrium outcome distributions across all games. This extends the comparison problem of \cite{Blackwell1951} from decision problems to games and from single experiments to generators of experiments, and connects it to Aumann's theory of common knowledge through CKCs and information loops.

The main conceptual object is the information loop. Local refinement inside each CKC is necessary for stochastic dominance, but it is not sufficient once the oracle's partition connects several CKCs.
Loops create global measurability constraints: the local experiments that mimic another oracle inside different CKCs must be glued into a single oracle-measurable public experiment. Covers and order-preserving covers describe when these loop constraints can be reproduced by the other oracle.

Using this geometry, we characterize oracle equivalence (mutual dominance) via two-sided refinement and order-preserving covers of irreducible loops (Theorem \ref{Theorem: Equivalent oracles}). For one-sided stochastic dominance, we prove that dominance implies refinement within each CKC and that every irreducible loop of the dominating oracle must admit a cover by the dominated oracle; when the relevant cover is unique, it must preserve cyclic order (Theorem \ref{Theorem: stochastic loop cover dominance}). Conversely, refinement together with order-preserving covers of irreducible $F_1$-loops is sufficient under separated-loop structures. The benchmark cases then follow naturally: with a unique CKC, or with no $F_1$-loops, dominance reduces to refinement inside CKCs (Corollary~\ref{Corollary: stochastic benchmark cases}); non-informative loops and the two-CKC case provide additional tractable environments.

\heading{Open issues.}
A full one-sided characterization remains open beyond the partial characterization and benchmark cases, and this also relates to challenging problems in graph theory.

Extending the comparison to general stochastic information structures, in which Markov kernels generate both the oracles' and the players' primitive information, raises further challenges. In particular, such an extension would require an analogue of information loops for stochastic information environments. We leave this for future research.

As noted earlier, following Definition \ref{Definition - Strategic Dominance}, there may be alternative ways to compare oracles beyond the one adopted here. It would be interesting to understand the logical relationship between these notions of comparison, if any exists.

\bibliographystyle{chicago}

\bibliography{refs.bib}

% ---------------------------------------

\appendix

%\begin{spacing}{1.1}

\section{Proofs and supplementary results} \label{Appendix: proofs}

Appendix \ref{Appendix: proofs} is organized as follows. 
We first prove the CKC-posterior characterization of dominance. We then derive local refinement inside a single CKC and the graph-theoretic loop constraints. The constructive sufficiency tools are collected afterward.

\subsection{Posterior characterization and local refinement} \label{Appendix: posterior mimicry and local refinement}

We first prove Proposition~\ref{Prop - Charact in terms of signaling funct}.

\begin{proof}
\textbf{If direction.}
Recall that \(Q_\tau(C)=\{P_{\tau,s,C}:\mathbb P_\tau(s\mid C)>0\}\). By construction, \(\overline{\mu}_{\tau_1\mid C}=\overline{\mu}_{\tau_2\mid C}\) if and only if the two experiments induce the same distribution over the posteriors \(P_{\tau_1,t,C}\) and \(P_{\tau_2,s,C}\).

Assume first that for every experiment $\tau_2$ there exists an experiment $\tau_1$ such that $\overline{\mu}_{\tau_1\mid C}=\overline{\mu}_{\tau_2\mid C}$ for every CKC $C$. Fix an experiment $\tau_2$ with signal space $S_2$, and let $\tau_1$ be an experiment of Oracle~$1$, with signal space $S_1$, such that $\overline{\mu}_{\tau_1\mid C}=\overline{\mu}_{\tau_2\mid C}$ for every CKC $C$. We construct below an augmented experiment $\hat\tau_1$ that depends only on $\tau_2$ and $\tau_1$. Now fix an arbitrary game $G$ and an equilibrium $(\sigma_i)_i$ of $G(\tau_2)$.

For each pair $(q,C)$ with $q\in Q_{\tau_2}(C)$, let $R_2(q,C):=\{s\in S_2:P_{\tau_2,s,C}=q\}$. Thus, $R_2(q,C)$ is the set of $\tau_2$-signals that induce the same posterior on $C$, and hence the same joint posterior profile within $C$. Different signals in $R_2(q,C)$ may correspond to different public labels and therefore may support different coordination behavior. So the augmented Oracle 1 experiment needs to randomly choose one of these Oracle 2 signal labels and match its distribution. For $s\in R_2(q,C)$, let $\kappa_{q,C}(s):=\tfrac{\mathbb P_{\tau_2}(s\mid C)}{\sum_{s'\in R_2(q,C)}\mathbb P_{\tau_2}(s'\mid C)}$ be the conditional distribution of the original $\tau_2$-signal label, given that the realized CKC is $C$ and the induced CKC-posterior is $q$. Let $\mathcal I$ be the finite set of such pairs $(q,C)$ that may arise under $\tau_2$. We introduce auxiliary common randomness by considering a selection rule.  

Let $R= \left\{ r:\mathcal I\to S_2: r(q,C)\in R_2(q,C), \ \forall (q,C)\in\mathcal I \right\}$. Thus $r$ specifies, for every possible pair $(q,C)$, a signal label $r(q,C)\in R_2(q,C)$. Equip $R$ with the product probability $\rho$ given by \(\rho(r) = \prod_{(q,C)\in\mathcal I} \kappa_{q,C}\bigl(r(q,C)\bigr)\). Hence the coordinate $r(q,C)$ is distributed according to $\kappa_{q,C}$, independently
across different pairs $(q,C)$. In the augmented experiment, only the coordinate
corresponding to the realized pair $(q,C)$ is used. This coordinate reproduces the
conditional distribution of the original $\tau_2$-signal labels among all signals that
induce the same posterior $q$ in $C$. Define the augmented signal space $\hat S_1:=S_1\times R$ and the experiment $\hat\tau_1$ by $\hat\tau_1((t,r)\mid\omega):=\tau_1(t\mid\omega)\rho(r)$. This preserves $F_1$-measurability since the auxiliary randomization is independent of the state.

We now define strategies under $\hat\tau_1$. Fix $\omega$, let $C$ be the CKC containing $\omega$, and suppose the realized signal is $(t,r)$. If $\mathbb P_{\tau_1}(t\mid C)>0$, put $q:=P_{\tau_1,t,C}$. By $\overline{\mu}_{\tau_1\mid C}=\overline{\mu}_{\tau_2\mid C}$, we have $q\in Q_{\tau_2}(C)$, so $r(q,C)$ is well defined. Set $\hat\sigma_i(a_i\mid\Pi_i(\omega),(t,r)) := \sigma_i(a_i\mid\Pi_i(\omega),r(q,C))$. Conditional on each CKC $C$ and posterior $q$, the state distribution and all players' posteriors under $\tau_1$ after $t$ are the same as under $\tau_2$ after any signal in $R_2(q,C)$. The auxiliary coordinate selects among those strategically distinct $\tau_2$-signals with the same conditional probabilities as $\tau_2$. Hence each player faces the same best-response problem under $\hat\tau_1$ as under $\tau_2$, so $(\hat\sigma_i)_i$ is an equilibrium of $G(\hat\tau_1)$ and induces the same state--action distribution as $(\sigma_i)_i$ under $\tau_2$. Therefore $\mathrm{NED}(G(\tau_2))\subseteq\mathrm{NED}(G(\hat\tau_1))$. Since $\hat\tau_1$ was chosen before fixing $G$ and the equilibrium, the same experiment $\hat\tau_1$ works for every game $G$.

\textbf{Only if direction.} Conversely, suppose the condition in the proposition fails. Then there exists an experiment $\tau_2^0$ of Oracle~$2$, with finite signal space $S_2$, such that for every experiment $\tau_1$ of Oracle~$1$, there is a CKC $C$ for which $\overline{\mu}_{\tau_1\mid C}\neq\overline{\mu}_{\tau_2^0\mid C}$. We may assume that the posteriors induced by this experiment have full support on each CKC. Indeed, choose $\lambda\in\Delta(S_2)$ with $\lambda(s)>0$ for every $s$, and define $\tau_2^\varepsilon(s\mid\omega):=(1-\varepsilon)\tau_2^0(s\mid\omega)+\varepsilon\lambda(s)$. This perturbation preserves measurability with respect to Oracle~$2$ and, assuming the prior has full support on each CKC, makes every posterior $P_{\tau_2^\varepsilon,s,C}$ have full support on $C$. If every sufficiently small $\tau_2^\varepsilon$ were mimicked by some Oracle~$1$ experiment, then compactness and continuity would yield an Oracle~$1$ experiment mimicking $\tau_2^0$, contradicting the choice of $\tau_2^0$.\footnote{Any mimicking signal has, in each CKC, either zero conditional probability or a posterior in $Q_{\tau_2^\varepsilon}(C)$. Hence there are at most $K=(|S_2|+1)^{|\mathcal C|}$ support--posterior profiles. Signals with the same profile can be merged, since their likelihood
vectors are proportional in every supported CKC, and merging preserves both $F_1$-measurability and every CKC posterior law. After padding with null signals, mimickers lie in the compact set of $F_1$-measurable experiments on a fixed $K$-signal space, and continuity yields the required limit.} Hence, for some $\varepsilon>0$, $\tau_2^\varepsilon$ still cannot be mimicked by Oracle~$1$. Rename this experiment $\tau_2$.

To disprove dominance, it suffices to show that, for every experiment $\tau_1$ of Oracle~$1$, some finite game has an equilibrium outcome under $\tau_2$ that $\tau_1$ cannot reproduce. Fix an arbitrary experiment $\tau_1$ of Oracle~$1$. For every player $i$ and CKC $C$, let $\mathcal P_i(C)\subseteq\Delta(C)$ be the finite set of private posteriors of player $i$ that arise with positive probability in $C$ under either $\tau_1$ or $\tau_2$.

We construct a finite game \(G\), which may depend on \(\tau_1\). Identify each distribution on a CKC with its zero extension to \(\Omega\), and give each player the fixed action set \(A_i=\left(\bigcup_C Q_{\tau_2}(C)\right) \times \left(\bigcup_C\mathcal P_i(C)\right)\), where the unions are over CKCs. Thus an action is a pair \((q_i,p_i)\) of distributions on \(\Omega\). At state \(\omega\), let \(C\) be the CKC containing \(\omega\). Let \(S_C\) be a bounded strictly proper scoring rule on \(\Delta(C)\), normalized so that \(S_C(p,\omega)>0\) for every \(p\in\Delta(C)\) and \(\omega\in C\). If all players report the same \(q_i\in Q_{\tau_2}(C)\) and player \(i\)'s second report satisfies \(p_i\in\mathcal P_i(C)\), player \(i\) receives \(S_C(p_i,\omega)\); otherwise, she gets \(0\). 

In $G(\tau_2)$, consider the equilibrium in which, after signal $s$ in CKC $C$, all players report $q_i=P_{\tau_2,s,C}$, and player $i$ reports $p_i=P_{\tau_2,s,C}(\cdot\mid\Pi_i(\omega)\cap C)$. This is an equilibrium: given that other players report the same $q$, strict propriety makes the truthful $p_i$-report optimal, and changing the $q_i$-coordinate breaks coordination and gives payoff $0$. Suppose that some equilibrium of $G(\tau_1)$ induces the same state--joint-action distribution. We show that this implies $\overline{\mu}_{\tau_1\mid C}=\overline{\mu}_{\tau_2\mid C}$ for every CKC $C$, contradicting the choice of $\tau_2$.

Fix a CKC \(C\). Since the selected equilibrium under \(\tau_2\) always has a common \(q\)-report, equality of state--joint-action distributions implies the same almost surely under the mimicking equilibrium of \(\tau_1\). Fix a signal \(t\) with \(\mathbb P_{\tau_1}(t\mid C)>0\), and choose \(\omega\in C\) such that \(\tau_1(t\mid\omega)>0\). Conditional on \((\omega,t)\), players' reports are independently drawn, so diagonal support forces all \(q\)-report marginals to be degenerate at a common \(q\in Q_{\tau_2}(C)\). For every player \(i\), strict propriety makes her second report equal her posterior after \((\Pi_i(\omega),t)\), while outcome matching makes this posterior equal to \(q(\cdot\mid\Pi_i(\omega)\cap C)\). Since \(q\) has full support on \(C\) and the prior has full support, \(\tau_1(t\mid\omega')>0\) for every \(\omega'\in\Pi_i(\omega)\cap C\). Player \(i\)'s \(q\)-report is therefore \(q\) throughout this cell, and diagonal support forces all players to report \(q\) at every newly reached state. Iterating along information-cell paths in \(C\) shows that \(t\) is positive throughout \(C\) and that the common \(q\)-report is constant. Denote it by \(g_C(t)\in Q_{\tau_2}(C)\). Equality of state--joint-action distributions then implies that $\mathbb P_{\tau_1}(g_C(t)=q\mid C) =\mathbb P_{\tau_2}(P_{\tau_2,s,C}=q\mid C)$ for every \(q\in Q_{\tau_2}(C)\). We claim that if \(g_C(t)=q\), then \(P_{\tau_1,t,C}=q\). Put \(r_t:= P_{\tau_1,t,C}\).

Suppose \(g_C(t)=q\). By strict propriety, after \((\Pi_i(\omega),t)\), player \(i\)'s \(p_i\)-report is degenerate at $r_t(\cdot\mid \Pi_i(\omega)\cap C)$. Under the selected equilibrium of \(G(\tau_2)\), conditional on state \(\omega\) and the common \(q\)-report, the corresponding report is degenerate at $q(\cdot\mid \Pi_i(\omega)\cap C)$. Under \(\tau_1\), this conditional report distribution is a mixture over signals \(t'\) satisfying \(g_C(t')=q\). Since it must equal the same point mass, every component receiving positive conditional weight must induce that report. The preceding paragraph shows that \(\tau_1(t\mid\omega)>0\) for every \(\omega\in C\). Therefore, for every player \(i\) and every \(\omega\in C\), $r_t(\cdot\mid\Pi_i(\omega)\cap C) = q(\cdot\mid\Pi_i(\omega)\cap C)$. Since \(q\) has full support, \(r_t/q\) is constant on every information cell. CKC connectedness propagates this constancy throughout \(C\), and normalization gives \(r_t=q\).

Therefore, whenever $g_C(t)=q$, the posterior induced by $\tau_1$ and $t$ equals $q$. Combining this with the preceding probability equality, for every $q\in Q_{\tau_2}(C)$ we get $\mathbb P_{\tau_1}(P_{\tau_1,t,C}=q\mid C)=\mathbb P_{\tau_2}(P_{\tau_2,s,C}=q\mid C)$. Thus, the two experiments induce the same distribution over posteriors in $C$, and therefore $\overline{\mu}_{\tau_1\mid C}=\overline{\mu}_{\tau_2\mid C}$. Since $C$ was arbitrary, this holds for every CKC $C$, contradicting the choice of $\tau_2$.  Thus, for the arbitrary experiment $\tau_1$ fixed above, the finite game $G$ has an equilibrium outcome under $\tau_2$ that $\tau_1$ cannot reproduce. Since $\tau_1$ was arbitrary, no experiment of Oracle~$1$ can reproduce every equilibrium outcome of $\tau_2$ in every game, so Oracle~$1$ does not dominate Oracle~$2$. This completes the proof.
\hfill
\end{proof}

The following observation is used only for necessity statements.

\begin{obs}[CKC-level proportionality] \label{Obs:CKC proportionality}
Fix a CKC \(C\), experiments \(\tau_1,\tau_2\), and signals \(t,s\) that occur with positive probability conditional on \(C\). If \(t\) under \(\tau_1\) and \(s\) under \(\tau_2\) induce the same posterior on \(C\), then there is \(c_{t,s,C}>0\) such that \(\tau_1(t|\omega)= c_{t,s,C} \tau_2(s|\omega)\) for every \(\omega\in C\).
\end{obs}

Indeed, equality of posteriors on \(C\) means that the two likelihood vectors are proportional on \(C\) by Bayes' rule and the full-support prior. The last supporting lemma states that refinement within every CKC is a necessary condition for dominance.

\begin{lemma}[Single-CKC refinement] \label{Lemma - single CKC refinement}
Assume that $C$ is a CKC.
The following are equivalent:
\begin{enumerate}[(i)]
    \item $F_1|_C$ refines $F_2|_C$;
    \item for every $F_2|_C$-measurable experiment $\tau_2$, there exists an
    $F_1|_C$-measurable experiment $\tau_1$ such that $\overline{\mu}_{\tau_1\mid C}=\overline{\mu}_{\tau_2\mid C}$.
\end{enumerate}
Consequently, if $F_1\succeq_{\rm NE}F_2$, then $F_1$ refines $F_2$ in every
CKC.
\end{lemma}

\begin{proof}
If $F_1|_C$ refines $F_2|_C$, every $F_2|_C$-measurable experiment is already $F_1|_C$-measurable, so the claim is immediate. Conversely, suppose $F_1|_C$ does not refine $F_2|_C$. Then there are $\omega,\bar\omega\in C$ with $F_1(\omega)=F_1(\bar\omega)$ but $F_2(\omega)\neq F_2(\bar\omega)$. Let \(A=F_2(\omega)\cap C\), and let \(\tau_2\) be the binary \(F_2|_C\)-measurable experiment that reveals whether the state lies in \(A\). Under \(\tau_2\), the signal \(A\) induces the posterior \(\mu(\cdot|A)\) on \(C\). If an \(F_1|_C\)-measurable experiment \(\tau_1\) satisfied \(\overline{\mu}_{\tau_1\mid C}=\overline{\mu}_{\tau_2\mid C}\), then some signal \(t\) of \(\tau_1\) would induce \(\mu(\cdot|A)\) on \(C\). But \(F_1\)-measurability implies \(\tau_1(t|\omega)=\tau_1(t|\bar\omega)\), while \(\omega\in A\) and \(\bar\omega\notin A\). Hence \(t\) cannot induce a posterior supported on \(A\), a contradiction. The final statement follows from Proposition~\ref{Prop - Charact in terms of signaling funct}.
\end{proof}

\subsection{Loop algebra} \label{Appendix: loop algebra}

The cover condition has a useful algebraic characterization. The next definition is mainly a test for coverability: it asks whether every binary $F_{-i}$-measurable split of the loop has the same number of transitions in the two directions.

\begin{definition} \label{Definition - F_2 balanced loops}
    An $F_i$-loop $(\omega_1,\overline{\omega}_1, \omega_2,\overline{\omega}_2, \dots, \omega_m,\overline{\omega}_m)$ is $F_{-i}$-\emph{balanced} if for every pair of disjoint $F_{-i}$-measurable sets $A$ and $B$ such that $\cup_{j=1}^m\{\omega_j,\overline{\omega}_j\}\subseteq A\cup B$, it follows that
    \begin{equation}\label{eq: balanced}
    \#(A \to B):=\bigl|\{j: \omega_j\in A,\ \overline{\omega}_j\in B\}\bigr|
    =
    \bigl|\{j: \omega_j\in B,\  \overline{\omega}_j\in A\}\bigr|=:\#(B \to A).
    \end{equation}
\end{definition}

Figure~\ref{fig: F1 loop and non-balanced loop}(a) depicts an $F_1$-loop with three CKCs.
Panel~(b) shows a binary $F_2$-measurable split of this loop into $A=\{\omega_1, \omega_2, \omega_3\}$ and $B=\{\overline{\omega}_1, \overline{\omega}_2, \overline{\omega}_3\}$. Since $\#(A\to B)=3$ and $\#(B\to A)=0$, the loop is not $F_2$-balanced and therefore, by Proposition~\ref{proposition: balanced}, not $F_2$-covered.

\begin{figure}[th!]
    \centering
    \begin{minipage}{0.45\linewidth}
        \centering
        \begin{tikzpicture}[scale=0.8]

        % Draw the rectangles with labels
        \draw (1.5,0.5) -- (4.5,0.5) -- (4.5,1.5) -- (1.5,1.5) -- cycle;
        \node at (2,1) {$\o_1$};
        \node at (4,1) {$\overline{\o}_1$};

        \draw (4.5,-1) -- (5.5,-1) -- (5.5,-3.5) -- (4.5,-3.5) -- cycle;
        \node at (5,-1.5) {$\o_2$};
        \node at (5,-3) {$\overline{\o}_2$};

        \draw (0.5,-1) -- (1.5,-1) -- (1.5,-3.5) -- (0.5,-3.5) -- cycle;
        \node at (1,-3) {$\o_3$};
        \node at (1,-1.5) {$\overline{\o}_3$};

        % Correct the ellipses
        \draw[blue, thick, rotate around={-65:(4.45,-0.25)}] (4.45,-0.25) ellipse (2.0cm and 0.7cm); % Blue ellipse
        \node[blue] at (4.4,-0.2) {$F_1(\omega_2)$};
        \draw[red, thick, rotate around={0:(3,-3)}] (3,-3) ellipse (2.4cm and 0.7cm); % Red ellipse
        \node[red] at (3,-3) {$F_1(\omega_3)$};
        \draw[teal, thick, rotate around={65:(1.5,-0.25)}] (1.5,-0.25) ellipse (2.0cm and 0.7cm); % Green ellipse
        \node[teal] at (1.5,-0.2) {$F_1(\omega_1)$};

        % Add labels
        \node at (3,1.8) {$C_1$};
        \node at (6,-2.5) {$C_2$};
        \node at (-0.1,-2.5) {$C_3$};

        \end{tikzpicture}
        \caption*{(a)}
    \end{minipage}    %  \hspace{1cm}
\begin{minipage}{0.45\linewidth}
    \centering
    \begin{tikzpicture}[scale=0.8]

    % Draw the rectangles with labels
    \draw (1.5,0.5) -- (4.5,0.5) -- (4.5,1.5) -- (1.5,1.5) -- cycle;
    \node at (2,1) {$\o_1$};
    \node at (4,1) {$\overline{\o}_1$};

    \draw (4.5,-1) -- (5.5,-1) -- (5.5,-3.5) -- (4.5,-3.5) -- cycle;
    \node at (5,-1.5) {$\o_2$};
    \node at (5,-3) {$\overline{\o}_2$};

    \draw (0.5,-1) -- (1.5,-1) -- (1.5,-3.5) -- (0.5,-3.5) -- cycle;
    \node at (1,-3) {$\o_3$};
    \node at (1,-1.5) {$\overline{\o}_3$};

    % Draw the red Bézier curve surrounding \omega states
    \draw[red, thick] (1.2,0)
    .. controls (1.5,2) and (2,2) .. (3,1)
    .. controls (5,-1) and (5,-1) .. (5.5,-1.4)
    .. controls (5.7,-1.9) and (6,-2.0) .. (0.8,-3.6)
    .. controls (0.3,-3.7) and (0.9,-2.3) .. (1.2,-2)
    .. controls (1.8,-1.3) and (2.0,-1.1) .. (1.6,-0.9)
    .. controls (1.4,-0.7) and (1.1,-0.5) .. (1.2,0);
    \node[red] at (0,0.5) {$A=F_2(\omega_1)$};

    % Draw the blue Bézier curve surrounding \overline{\omega} states
    \draw[blue, thick] (4.8,0)
    .. controls (4.5,2) and (4,2) .. (3,1)
    .. controls (1,-1) and (1,-1) .. (0.5,-1.4)
    .. controls (0.3,-1.9) and (0,-2.0) .. (5.2,-3.6)
    .. controls (5.7,-3.7) and (5.1,-2.3) .. (4.8,-2)
    .. controls (4.2,-1.3) and (4,-1.1) .. (4.4,-0.9)
    .. controls (4.6,-0.7) and (4.9,-0.5) .. (4.8,0);
    \node[blue] at (6,0.5) {$B=F_2(\overline{\omega}_1)$};

    % Add labels
    \node at (3,1.8) {$C_1$};
    \node at (6,-2.5) {$C_2$};
    \node at (-0.1,-2.5) {$C_3$};

    \end{tikzpicture}
    \caption*{(b)}
\end{minipage}
     \caption{ \footnotesize  Figure (a) depicts an $F_1$-loop with three CKCs and six states overall. Figure (b) illustrates how the irreducible $F_1$-loop, presented in $(a)$,  is non-balanced with respect to $F_2$. Namely, $F_2$ has two elements $A=\{\omega_1,\omega_2,\omega_3\}$, and $B=\{\overline{\omega}_1, \overline{\omega}_2, \overline{\omega}_3\}$ such that the number of transitions from $A$ to $B$ are $3$, while the reverse equals $0$. See Appendix~\ref{Appendix: balanced-cover motivation} for a discussion on this figure and motivation.}
    \label{fig: F1 loop and non-balanced loop}
\end{figure}

\begin{proposition}[Cover criterion] \label{proposition: balanced}
   Let $(\omega_1,\overline{\omega}_1, \omega_2,\overline{\omega}_2 , \dots, \omega_m,\overline{\omega}_m)$ be an irreducible $F_1$-loop. The following statements are equivalent:
      \begin{description}
                 \item [i.] The loop is $F_2$-balanced;
                 \item [ii.] The loop is $F_2$-covered;
                 \item [iii.] For every $F_2$-measurable function $f:\{\omega_1,\overline{\omega}_1, \omega_2,\overline{\omega}_2 , \dots, \omega_m,\overline{\omega}_m\}\to (0,\infty)$, it follows that $\prod_{j=1}^m \tfrac{f(\omega_j)}{f(\overline{\omega}_j)}=1$.
      \end{description}
\end{proposition}

For irreducible loops, the third part of Proposition~\ref{proposition: balanced}, combined with Eq.~\eqref{eq: loop-product-constraint}, shows that covers are exactly the configurations under which every $F_2$-measurable likelihood-ratio product is neutral around the loop, and is therefore necessary for dominance.

\begin{proof}
{\bf iii} $\Rightarrow$ {\bf i}.  Suppose that $L=(\omega_1,\bar\omega_1,\ldots,\omega_m,\bar\omega_m)$ is not $F_2$-balanced.  Then there is an $F_2$-measurable partition $\{A,B\}$ of the states of the loop such that $\#(A\to B)\neq \#(B\to A)$.  Define $f(\omega)=\mathbf 1_{\{\omega\in A\}}+2\cdot \mathbf 1_{\{\omega\in B\}}$. Then $\prod_{j=1}^m \tfrac{f(\omega_j)}{f(\bar\omega_j)} =\tfrac12^{\#(A\to B)}2^{\#(B\to A)}\neq 1$, contradicting {\bf iii}. 

{\bf i} $\Rightarrow$ {\bf ii}.  Assume that $L$ is $F_2$-balanced.  For any $F_2$-atom $D$ that meets the loop, apply balancedness to the partition $\{D,D^c\}$.  Since $|\{j:\omega_j\in D\}|=\#(D\to D)+\#(D\to D^c)$ and $|\{j:\bar\omega_j\in D\}|=\#(D^c\to D)+\#(D\to D)$, balancedness implies $|\{j:\omega_j\in D\}|=|\{j:\bar\omega_j\in D\}|$.  Call this condition atom-balance.

Let $J=\{j:\omega_j\in F_2(\bar\omega_j)\}$.  These indices are already covered by the trivial part of the cover. 
Remove them from the list.  Since each removed pair has both states in the same $F_2$-atom, atom-balance continues to hold for the remaining pairs.

If no pairs remain, we are done.  Otherwise choose one remaining pair $(\omega_{j_1},\bar\omega_{j_1})$.  Since $j_1\notin J$, $\omega_{j_1}\notin F_2(\bar\omega_{j_1})$.  By atom-balance, the atom $F_2(\bar\omega_{j_1})$ contains some state $\omega_{j_2}$ from a remaining pair.  Thus $\omega_{j_2}\in F_2(\bar\omega_{j_1})$.  Repeating the same argument gives a sequence $j_1,j_2,\ldots$ such that $\omega_{j_{r+1}}\in F_2(\bar\omega_{j_r})$ for every $r$.  Since there are finitely many remaining indices, some index eventually repeats.  The first closed segment obtained in this way is an $F_2$-loop. Indeed, since the original loop is irreducible, no CKC appears more than once in it; hence the closed segment satisfies the distinct-CKC condition in Definition~\ref{Definition: a loop}. All selected states are distinct, and the remaining conditions follow from the construction.

Remove the indices of this $F_2$-loop. This preserves atom-balance, because from each $F_2$-atom it removes the same number of $\omega$-states and $\bar\omega$-states. Repeating the procedure decomposes all remaining pairs into $F_2$-loops.  Together with $J$, this gives an $F_2$-cover of $L$.

{\bf ii} $\Rightarrow$ {\bf iii}. 
Assume that $L$ is $F_2$-covered. 
Then $\{1,\ldots,m\}$ is partitioned into $J,I_1,\ldots,I_r$, where $\omega_j\in F_2(\bar\omega_j)$ for every $j\in J$ and each $I_t$ forms an $F_2$-loop. 
Let $f$ be positive and $F_2$-measurable. 
If $j\in J$, then $f(\omega_j)=f(\bar\omega_j)$, so $\tfrac{f(\omega_j)}{f(\bar\omega_j)}=1$. 
If $I_t$ is an $F_2$-loop, then $\prod_{j\in I_t} \tfrac{f(\omega_j)}{f(\bar\omega_j)}=1$, because the denominator at each step equals the numerator at the next step around the loop. 
Hence $\prod_{j=1}^m \tfrac{f(\omega_j)}{f(\bar\omega_j)}=1$. 
This proves {\bf iii}.
\end{proof}

The next Proposition~\ref{Proposition - irreducible and informative loops} records elementary facts about irreducible and informative loops. 

\begin{proposition} \label{Proposition - irreducible and informative loops}
Consider an $F_i$-loop $L_i$.
\begin{itemize}
    \item If $L_i$ intersects the same \emph{CKC} more than once, then it is not irreducible.
    \item If $L_i$ is irreducible and consists of at least $6$ states, then it is $F_i$-fully informative.
    \item If $L_i$ is not irreducible, then either it intersects the same \emph{CKC} more than once, or it has at least $4$ states in the same partition element of $F_i$.
\end{itemize}
\end{proposition}

\begin{proof}
We prove the three claims in order.

First, suppose $L_i$ contains two non-adjacent pairs $(\omega_a,\bar\omega_a)$ and $(\omega_b,\bar\omega_b)$ in the same CKC.  
By switching the partners inside that CKC, the two outside arcs between the pairs close to two shorter $F_i$-loops. The resulting sequences satisfy Definition~\ref{Definition: a loop}: their connecting sets are inherited from the original loop, and the newly paired states are distinct. Hence, the original loop is not irreducible.

Second, suppose $L_i$ is irreducible, has at least three CKC-pairs, and contains
a non-informative pair $(\omega_\ell,\bar\omega_\ell)$ with $F_i(\omega_\ell) = F_i(\bar\omega_\ell)$.  
If deleting this pair from the loop creates two neighboring pairs in the same CKC, the first paragraph applies.  
Otherwise, the remaining pairs still form an $F_i$-loop, contradicting irreducibility.
Thus an irreducible loop with at least six states is $F_i$-fully informative.

Finally, assume that \(L_i\) is not irreducible and does not intersect any CKC more than once. Suppose, toward a contradiction, that no \(F_i\)-atom contains four states of \(L_i\). Each original connecting pair \(\{\bar\omega_j,\omega_{j+1}\}\) where \(j=1,\ldots,m\), is contained in one \(F_i\)-atom. Because all states of the loop are distinct, an atom meeting two different connecting pairs would contain both pairs and hence at least four states. Thus the original connecting pairs lie in distinct \(F_i\)-atoms, and every \(F_i\)-connection between states of \(L_i\) is one of these original connections. Since the CKC-pairs are also distinct, the CKC-pairs and the original \(F_i\)-connections form a simple alternating cycle, which contains no strict sub-cycle. This contradicts the existence of a strict sub-loop.
\end{proof}

The following graph decomposition is used in the proof of Theorem~\ref{Theorem: Equivalent oracles}.

\begin{lemma}[Atom-block decomposition] \label{Lemma - atom block decomposition}
For a partition \(F\), let \(\Gamma_F\) be the bipartite incidence graph whose vertices are the CKCs and the \(F\)-atoms, with \(C\) adjacent to \(A\) if \(C\cap A\neq\emptyset\). Define \(C\sim_F C'\) if \(C\) and \(C'\) remain connected in \(\Gamma_F\setminus\{A\}\) for every \(F\)-atom \(A\). Then \(\sim_F\) is an equivalence relation. Moreover:
\begin{enumerate}[(i)]
    \item if \(C\neq C'\), \(C\sim_F C'\), and an \(F\)-atom \(A\) meets both \(C\) and \(C'\), then any prescribed states \(\omega\in A\cap C\) and \(\omega'\in A\cap C'\) belong to a fully informative irreducible \(F\)-loop containing the designated \(F\)-connection between \(\omega\) and \(\omega'\);
    \item the bipartite graph whose vertices are the \(\sim_F\)-classes and the \(F\)-atoms meeting more than one class, with adjacency defined by intersection, is a forest.
\end{enumerate}
\end{lemma}

\begin{proof}
Reflexivity and symmetry are immediate. For transitivity, fix an \(F\)-atom \(A\): paths from \(C\) to \(C'\) and from \(C'\) to \(C''\) in \(\Gamma_F\setminus\{A\}\) concatenate to a path from \(C\) to \(C''\).

For (i), choose a shortest path from \(C\) to \(C'\) in \(\Gamma_F\setminus\{A\}\) and add the two incidences \(C-A-C'\). This gives a simple cycle. Selecting one state from each incidence, using the prescribed states at \(A\), produces a fully informative irreducible \(F\)-loop satisfying Definition~\ref{Definition: a loop} and containing the designated connection.

For (ii), suppose the stated quotient graph contains a cycle \(B_1,A_1,B_2,A_2,\ldots,B_k,A_k,B_1\). Choose CKCs \(C_1\in B_1\) and \(C_2\in B_2\) met by \(A_1\). If \(D\neq A_1\), the path \(C_1-A_1-C_2\) survives in \(\Gamma_F\setminus\{D\}\). If \(D=A_1\), follow the remainder of the quotient cycle and, inside each class \(B_j\), join the relevant CKCs by a path in \(\Gamma_F\setminus\{A_1\}\). Thus \(C_1\) and \(C_2\) remain connected after deletion of every \(F\)-atom, contradicting \(B_1\neq B_2\).
\end{proof}

\subsection{Necessity of loop covers}
\label{Appendix: necessity of loop covers}

\begin{lemma}[Cover necessity] \label{Lemma - cover necessity}
If $F_1\succeq_{\rm NE}F_2$, then every irreducible $F_1$-loop is $F_2$-covered.
\end{lemma}

\begin{proof}
Suppose, to the contrary, that an irreducible $F_1$-loop $(\omega_1,\overline{\omega}_1, \dots, \omega_m,\overline{\omega}_m)$ is not $F_2$-covered. By Proposition~\ref{proposition: balanced}, it is not $F_2$-balanced. This means that there is an $F_2$-measurable partition $\{A,B\}$ of these states such that Eq.\ \eqref{eq: balanced} is not satisfied. Define an $F_2$-measurable experiment that obtains two signals, $\a$ and $\b$, 
\begin{equation}\label{eq: D}
\tau_2(\a | \o) =\begin{cases}
      x, & \text{if } \o \in A , \\
       y, & \text{if }  \o \in B,
    \end{cases}
\end{equation}
and $\tau_2(\b | \omega)=1-\tau_2(\a | \omega)$.
On other states (outside the loop), $\tau_2$ is defined arbitrarily. The numbers $x,y\in (0,1)$ are chosen so that $\frac{\ln{x}-\ln{y}}{\ln{(1-x)}-\ln{(1-y)}}$ is irrational. By dominance and Proposition~\ref{Prop - Charact in terms of signaling funct}, there is an \(F_1\)-measurable experiment \(\tau_1\) such that \(\overline{\mu}_{\tau_1\mid C}=\overline{\mu}_{\tau_2\mid C}\) for every CKC \(C\). The following claims are straightforward.

\heading{Claim 1:} In every CKC \(C\), any signal of \(\tau_1\) that has positive probability conditional on \(C\) induces the same CKC-posterior as either \(\a\) or \(\b\) under \(\tau_2\).

\heading{Claim 2:} For any signal $s$ of $\tau_1$ that has positive probability at some state of the loop, and hence at every loop state, and for any $i$, $\frac{\tau_1(s|\o_i)}{\tau_1(s|\overline{\omega}_i)} \in \{\frac{x}{y},\frac{1-x}{1-y},\frac{y}{x},  \frac{1-y}{1-x},1\}$. Therefore,
$$\prod_{i=1}^m  \frac{\tau_1(s|\o_i)}{\tau_1(s|\overline{\omega}_i)}=
\left(\frac{x}{y}\right)^{\l_1}\cdot
\left(\frac{1-x}{1-y}\right)^{\l_2}\cdot
\left(\frac{y}{x}\right)^{k_1}\cdot
\left(\frac{1-y}{1-x}\right)^{k_2},$$
where $\l_1+\l_2= |\{i; \o_i\in A \ {\rm{and}} \ \overline{\omega}_i\in B\}|$ and
$k_1+k_2= |\{i; \o_i\in B \ {\rm{and}} \ \overline{\omega}_i\in A\}|$.

\heading{Claim 3:} For any signal $s$ of $\tau_1$ that has positive probability at some state of the loop, and hence at every loop state, $\prod_{i=1}^m  \frac{\tau_1(s|\o_i)}{\tau_1(s|\overline{\omega}_i)}=1$.

We therefore obtain $(\frac{x}{y})^{\l_1}(\frac{1-x}{1-y})^{\l_2}(\frac{y}{x})^{k_1}(\frac{1-y}{1-x})^{k_2}=1$. We conclude that there are whole numbers, say $\l=\l_1-k_1$ and $ k=k_2-\l_2$ such that $(\frac{x}{y})^{\l }=(\frac{1-x}{1-y})^{k}$.
Since $\frac{\ln{x}-\ln{y}}{\ln{(1-x)}-\ln{(1-y)}}=\frac{\ln{\frac{x}{y}}}{\ln{\frac{1-x}{1-y}}}$ is irrational, $\l=k=0$, implying that Eq.\ (\ref{eq: balanced}) is satisfied.
A contradiction.
\end{proof}

\begin{lemma}[Unique covers are order-preserving] \label{Lemma - unique cover order preserving}
Assume \(F_1\succeq_{\rm NE}F_2\). 
If an irreducible \(F_1\)-loop has a unique \(F_2\)-cover, then that cover is order-preserving.
\end{lemma}

\begin{proof}
Let \(L_1=(\omega_1,\bar\omega_1,\ldots,\omega_m,\bar\omega_m)\) be an irreducible \(F_1\)-loop with a unique \(F_2\)-cover, and let \(I=\{1,\ldots,m\}\). In a unique cover, every \(F_2\)-non-informative pair belongs to \(J\): otherwise it can be moved from its sub-loop to \(J\) (and, in the two-pair case, both pairs can be moved), producing another cover. For each \(i\in I\), write \(\alpha(i)=F_2(\omega_i)\) and \(\gamma(i)=F_2(\bar\omega_i)\), and form the directed atom multigraph \(H\) whose edge \(e_i\) goes from \(\alpha(i)\) to \(\gamma(i)\). 

Let \(H^\circ\) be obtained by deleting its self-loops.  Thus the self-loops of \(H\) are exactly the indices in \(J\). Each \(F_2\)-sub-loop gives a directed closed trail in \(H^\circ\). By Proposition~\ref{Proposition - irreducible and informative loops}, \(L_1\) visits every CKC at most once, so every directed simple cycle in a decomposition of such a trail is an admissible \(F_2\)-sub-loop: it inherits distinct states and CKC-pairs from \(L_1\), and its edge incidences give the required \(F_2\)-connections. Hence any alternative directed-cycle partition of \(H^\circ\), together with the same \(J\), would give a second \(F_2\)-cover. Therefore, the induced directed-cycle partition is unique.

Fix a weakly connected component \(K\) of \(H^\circ\). The cover supplies a directed-cycle partition of \(K\), and that partition is unique. Hence Theorem~2.8 of \cite{Cooper2025} applies: \(K\) is a bridgeless cactus digraph and its unique partition is the set \(B(K)\) of directed cycles. Moreover, the proof of Lemma~2.7 of \cite{Cooper2025} implies that, for every \(\beta\in B(K)\), each weakly connected component of \(K\setminus E(\beta)\) meets \(\beta\) in exactly one vertex.

Fix a directed cycle \(\beta\in B(K)\), and let \(U\subseteq I\) be the set of indices of the edges of \(\beta\). 
Suppose, toward a contradiction, that the corresponding \(F_2\)-sub-loop is not order-preserving relative to the cyclic order of \(L_1\). 
After cyclically renaming the relevant indices, write the selected three indices as \(1<j<k\) in the \(F_1\)-order, with \(\omega_k\) preceding \(\omega_j\) in the \(F_2\)-order of \(\beta\).

Partition the vertices of \(\beta\) into three directed arcs. Let \(V^2_1\) be the vertices encountered from \(\gamma(1)\) through \(\alpha(k)\), let \(V^2_k\) be those encountered from \(\gamma(k)\) through \(\alpha(j)\), and let \(V^2_j\) be those encountered from \(\gamma(j)\) through \(\alpha(1)\). These three sets are disjoint and partition \(V(\beta)\). Similarly, partition the states of \(L_1\) into the three \(F_1\)-arcs \(A^1_1\), from \(\bar\omega_1\) to \(\omega_j\); \(A^1_j\), from \(\bar\omega_j\) to \(\omega_k\); and \(A^1_k\), from \(\bar\omega_k\) to \(\omega_1\).

Set \(p_1=\frac15\), \(p_2=\frac13\), \(p_3=\frac23\), and \(p_4=\frac12\). Define a binary \(F_2\)-measurable experiment \(\tau_2\) with distinct signals \(s_1,s_2\) by setting \(\tau_2(s_1|\omega) =p_r\) whenever \(F_2(\omega)\in V^2_r\), for \(r\in\{1,k,j\}\), where \(p_k:=p_2\) and \(p_j:=p_3\). For each weak component \(C\) of \(K\setminus E(\beta)\), assign to every atom of \(C\) the same probability as its unique attachment atom on \(\beta\). On atoms outside \(K\), set \(\tau_2(s_1|\cdot)=p_4\). Finally set \(\tau_2(s_2|\cdot)=1-\tau_2(s_1|\cdot)\). Because the three vertex arcs partition \(V(\beta)\) and every attached component has a unique attachment vertex, this is well-defined and \(F_2\)-measurable. By construction, for every \(\ell\notin\{1,j,k\}\) and every \(s\in\{s_1,s_2\}\), we have \(\tau_2(s|\omega_\ell)=\tau_2(s|\bar\omega_\ell)\).

By dominance and Proposition~\ref{Prop - Charact in terms of signaling funct}, there is an \(F_1\)-measurable experiment \(\tau_1\) such that \(\overline{\mu}_{\tau_1\mid C}=\overline{\mu}_{\tau_2\mid C}\) for every CKC \(C\). Fix a signal \(t\) that is assigned positive probability at some state of \(L_1\). Signals assigning zero probability to every state of \(L_1\) do not enter the row-sum equations below. Since \(\tau_2\) has full-support posteriors, equality of \(\overline{\mu}_{\tau_1\mid C}\) and \(\overline{\mu}_{\tau_2\mid C}\), together with \(F_1\)-measurability along \(L_1\), implies that \(t\) is positive in every CKC of the loop. In each such CKC, the posterior induced by \(t\) under \(\tau_1\) is induced by either \(s_1\) or \(s_2\) under \(\tau_2\). By Observation~\ref{Obs:CKC proportionality}, the likelihood ratio of \(t\) under \(\tau_1\) in that CKC is equal to the likelihood ratio of either \(s_1\) or \(s_2\) under \(\tau_2\). Hence \(\tau_1(t|\omega_\ell)=\tau_1(t|\bar\omega_\ell)\) for every \(\ell\notin\{1,j,k\}\). Together with \(F_1\)-measurability along \(L_1\), this implies that, on the states of \(L_1\), the likelihood vector of \(t\) is constant on the three \(F_1\)-arcs: write these constants as \(a_t\) on \(A^1_1\), \(b_t\) on \(A^1_j\), and \(c_t\) on \(A^1_k\).

At the three exceptional CKCs, Observation~\ref{Obs:CKC proportionality} gives \(\tfrac{c_t}{a_t}\in \{\tfrac{p_3}{p_1}, \tfrac{1-p_3}{1-p_1}\}\), \(\tfrac{a_t}{b_t} \in \{\tfrac{p_2}{p_3},\tfrac{1-p_2}{1-p_3}\}\), and \(\tfrac{b_t}{c_t} \in \{ \tfrac{p_1}{p_2} , \tfrac{1-p_1}{1-p_2} \}\). Since \(\tfrac{c_t}{a_t}\cdot\tfrac{a_t}{b_t}\cdot\tfrac{b_t}{c_t}=1\), a direct calculation using \(p_1=\frac15\), \(p_2=\frac13\), and \(p_3=\frac23\) shows that the only possible choices are either $\left(\frac{p_3}{p_1}, \frac{p_2}{p_3}, \frac{p_1}{p_2}\right)$ or $\left(\frac{1-p_3}{1-p_1}, \frac{1-p_2}{1-p_3}, \frac{1-p_1}{1-p_2}\right)$.

Call signals of the first kind type \(1\), and signals of the second kind type \(2\).  Let \(x\) and \(y\) denote the total probability mass of types $1$ and $2$ at \(\omega_1\) respectively. The row-sum constraint at \(\bar\omega_1\) gives \(\frac{p_1}{p_3} x+\frac{1-p_1}{1-p_3}y=1\). Together with \(x+y=1\), this implies \(x=p_3\). The row-sum constraint at \(\bar\omega_j\) gives \(\frac{p_1}{p_2} x+\frac{1-p_1}{1-p_2}y=1\). Together with \(x+y=1\), this implies \(x=p_2\). This contradiction shows that \(\beta\) must be order-preserving. Since \(\beta\in B(K)\) was arbitrary, every directed cycle in the unique cycle partition of \(K\) is order-preserving.  Since \(K\) was arbitrary, the unique \(F_2\)-cover of \(L_1\) is order-preserving.
\end{proof}

\subsection{Constructive sufficiency tools} \label{Appendix: constructive sufficiency tools}

\begin{definition}[Posterior-equivalent extension of a binary experiment] \label{Definition - posterior equivalent extension}
Let \(a,b\in\Delta(S_1)\) be two probability laws on a finite signal set \(S_1\). 
Let \(q,q'\in\Delta(S_2)\) be two probability laws on another finite signal set \(S_2\). We say that the ordered pair \((q,q')\) is a \emph{posterior-equivalent extension} of \((a,b)\) if one of the following holds:

\begin{enumerate}
    \item There exist a finite set \(R\), a probability law \(\nu\in\Delta(R)\), and an identification \(S_2=S_1\times R\) such that \(q(s,r)=a(s)\nu(r)\) and \(q'(s,r)=b(s)\nu(r)\) for all \((s,r)\in S_1\times R\).
    \item \(a=b\) and \(q=q'\).
\end{enumerate}
In case (1), the coordinate \(r\) is common noise; in case (2), both experiments are uninformative.
\end{definition}

The next lemma constructs a mimicking experiment on a single loop with an order-preserving cover.

\begin{lemma}[Single-loop construction] \label{Lemma - single loop construction}
Let \(L=(\omega_1,\bar\omega_1,\ldots,\omega_m,\bar\omega_m)\) be an \(F_1\)-loop, with an order-preserving \(F_2\)-cover \(\{1,\ldots,m\}=J\cup I_1\cup\cdots\cup I_K\), where \(\omega_i\in F_2(\bar\omega_i)\) for every \(i\in J\). Let \(\tau_2\) be an \(F_2\)-measurable experiment on a finite set \(S_2\). Assume that the \(F_1\)-atoms \(F_1(\bar\omega_t)=F_1(\omega_{t+1})\), \(t=1,\ldots,m\), are distinct. Then there is an \(F_1\)-measurable experiment \(\tau_1\) on \(S_2^K\times S_2^J\) such that, for every \(i\), the ordered pair \((\tau_1(\cdot|\omega_i),\tau_1(\cdot|\bar\omega_i))\) is a posterior-equivalent extension of \((\tau_2(\cdot|\omega_i),\tau_2(\cdot|\bar\omega_i))\) satisfying case (1) of Definition~\ref{Definition - posterior equivalent extension}.
\end{lemma}

\begin{proof}
For each \(i\), set \(p_i:=\tau_2(\cdot|\bar\omega_i)\in\Delta(S_2)\). If \(i\in J\), then \(\omega_i\in F_2(\bar\omega_i)\), so \(\tau_2(\cdot|\omega_i)=p_i\). If \(i\in I_k\), let \({\rm pred}(i)\) be the predecessor of \(i\) in the \(F_2\)-sub-loop indexed by \(I_k\). Since the cover is order-preserving, \(\tau_2(\cdot|\omega_i)=p_{{\rm pred}(i)}\) and \(\tau_2(\cdot|\bar\omega_i)=p_i\).

For each block \(I_k\), write its elements in the cyclic order inherited from \(L\). For \(t=1,\ldots,m\), let \(\ell_k(t)\) be the last element of \(I_k\) weakly before \(t\) in that cyclic order, and if there is no such element before \(t\), let \(\ell_k(t)\) be the last element of \(I_k\). Define \(q_t\in\Delta(S_2^K\times S_2^J)\) by
\[
q_t((x_k)_{k=1}^K,(y_j)_{j\in J})
=
\prod_{k=1}^K p_{\ell_k(t)}(x_k)\prod_{j\in J}p_j(y_j).
\]
Now define \(\tau_1\) on the \(F_1\)-atoms in the loop by \(\tau_1(\cdot|\bar\omega_t)=\tau_1(\cdot|\omega_{t+1})=q_t\), with indices modulo \(m\). Since these \(F_1\)-atoms are distinct, this assignment is unambiguous. Extend \(\tau_1\) arbitrarily, but \(F_1\)-measurably, on \(F_1\)-atoms not appearing in the loop.

Fix \(i\). Under \(\tau_1\), the two laws in the CKC-pair \((\omega_i,\bar\omega_i)\) are \(q_{i-1}\) and \(q_i\). If \(i\in I_k\), then order preservation gives \(\ell_k(i-1)={\rm pred}(i)\) and \(\ell_k(i)=i\), while for every \(h\neq k\), \(\ell_h(i-1)=\ell_h(i)\). Hence \(q_{i-1}\) and \(q_i\) differ only in coordinate \(k\), whose laws are \(\tau_2(\cdot|\omega_i)\) and \(\tau_2(\cdot|\bar\omega_i)\), while all other coordinates are common noise. If \(i\in J\), then \(\ell_k(i-1)=\ell_k(i)\) for every \(k\), and \(\tau_2(\cdot|\omega_i) = \tau_2(\cdot|\bar\omega_i)=p_i\). Taking the \(J\)-coordinate indexed by \(i\) as the \(S_2\)-coordinate, all other coordinates are common noise. Either way, case (1) of Definition~\ref{Definition - posterior equivalent extension} holds.
\end{proof}

\subsection{Proof of Theorem~\ref{Theorem: Equivalent oracles}}

\begin{proof}
Assume first that $F_1$ and $F_2$ are equivalent. By Lemma~\ref{Lemma - single CKC refinement} and Lemma~\ref{Lemma - cover necessity}, applied in both directions, for each $i$, $F_i$ refines $F_{-i}$ in every CKC and every irreducible $F_i$-loop is $F_{-i}$-covered. Since refinement holds in both directions, $F_1|_C=F_2|_C$ for every CKC $C$.

Fix $i$, and let $L_i$ be a fully informative irreducible $F_i$-loop. No index in an $F_{-i}$-cover of $L_i$ can belong to $J$, because $F_i|_C=F_{-i}|_C$ in every CKC. If the cover contained a proper $F_{-i}$-sub-loop, a simple-cycle component of that sub-loop would be a fully informative irreducible $F_{-i}$-loop. By Lemma~\ref{Lemma - cover necessity}, applied in the reverse direction, it has an $F_i$-cover. Again local equality rules out non-informative pairs, so this cover contains a proper $F_i$-sub-loop of $L_i$, contradicting irreducibility. Hence every cover of $L_i$ consists of a single $F_{-i}$-loop using all its pairs. This cover is unique: otherwise the directed $F_{-i}$-atom graph would contain a proper directed cycle, and the preceding argument would again contradict irreducibility. Lemma~\ref{Lemma - unique cover order preserving} therefore implies that the cover preserves the cyclic order. Finally, if an irreducible $F_i$-loop is not $F_i$-fully informative, Proposition~\ref{Proposition - irreducible and informative loops} implies that it has exactly two CKC-pairs, so order preservation is automatic. This proves the stated conditions.

Conversely, assume the conditions in the theorem. We prove $F_1\succeq_{\rm NE}F_2$; the reverse direction is symmetric. Mutual local refinement gives $F_1|_C=F_2|_C$ for every CKC $C$.

We first show that every fully informative irreducible $F_1$-loop is an $F_2$-loop with the same cyclic order. Let $L$ be such a loop. Its order-preserving $F_2$-cover contains no non-informative pair. If it contained a proper $F_2$-sub-loop, a fully informative irreducible simple-cycle component of that sub-loop would, by the theorem's reverse hypothesis, have an $F_1$-cover. Local equality again rules out non-informative pairs, so this cover would contain a proper $F_1$-sub-loop of $L$, a contradiction. Hence the cover consists of a single $F_2$-loop, and it has the cyclic order of $L$.

Let $\Gamma_1:=\Gamma_{F_1}$ and let the \emph{clusters} be the $\sim_{F_1}$-classes from Lemma~\ref{Lemma - atom block decomposition}. We claim that if $\omega$ and $\omega'$ lie in CKCs of the same cluster and $F_1(\omega)=F_1(\omega')$, then $F_2(\omega)=F_2(\omega')$. If the two states lie in the same CKC, this follows from $F_1|_C=F_2|_C$. Otherwise, Lemma~\ref{Lemma - atom block decomposition}(i) gives a fully informative irreducible $F_1$-loop containing the designated $F_1$-connection between $\omega$ and $\omega'$. By the preceding paragraph, this is an $F_2$-loop with the same cyclic order, so $F_2(\omega)=F_2(\omega')$. Consequently, every $F_2$-measurable experiment restricted to a cluster is also $F_1$-measurable.

Call a union of clusters \emph{solvable} if every $F_2$-measurable experiment on that union can be mimicked there by an $F_1$-measurable experiment which, inside each CKC, reproduces the target experiment up to common noise. Each individual cluster is solvable. We use the following gluing step. Let $A_1,A_2$ be disjoint solvable unions of clusters and assume that at most one $F_1$-atom meets both. Fix an $F_2$-measurable experiment $\tau_2$ on $A_1\cup A_2$, and let $\tau_1^i$ be a mimicking experiment on $A_i$ with signal space $S_i$. If a connecting $F_1$-atom exists, let $\lambda_i$ be the law of $\tau_1^i$ on its intersection with $A_i$; otherwise choose any law on $S_i$. On $S_1\times S_2$, define
\[
\tau_1((s_1,s_2)\mid\omega)=
\begin{cases}
\tau_1^1(s_1\mid\omega)\lambda_2(s_2), & \omega\in A_1,\\
\lambda_1(s_1)\tau_1^2(s_2\mid\omega), & \omega\in A_2.
\end{cases}
\]
This experiment is $F_1$-measurable; on the possible connecting atom, both sides assign $\lambda_1\otimes\lambda_2$, and inside each $A_i$ the other coordinate is common noise. Hence $A_1\cup A_2$ is solvable.

We identify each cluster with the union of its CKCs. Let \(T\) be the bipartite forest in Lemma~\ref{Lemma - atom block decomposition}(ii). We prove solvability by induction on the number of cluster-vertices of \(T\). The one-cluster case follows because every individual cluster is
solvable. Otherwise, \(T\) has a cluster-vertex \(B\) of degree at most one: either \(B\) is isolated, or it is a leaf of a nontrivial component, since every atom-vertex has degree at least two. Let \(U\) be the union of all remaining clusters. After deleting \(B\) and any atom-vertices that no longer meet two remaining clusters, one obtains a forest with one fewer cluster-vertex, so \(U\) is solvable by induction. Any \(F_1\)-atom meeting both \(B\) and \(U\) is an atom-vertex adjacent to \(B\), so there is at most one such atom. The preceding gluing step therefore implies that \(B\cup U=\Omega\) is solvable.

Hence, for every $F_2$-measurable experiment $\tau_2$, there is an $F_1$-measurable experiment $\tau_1$ such that $\overline{\mu}_{\tau_1\mid C}=\overline{\mu}_{\tau_2\mid C}$ for every CKC $C$. Proposition~\ref{Prop - Charact in terms of signaling funct} gives $F_1\succeq_{\rm NE}F_2$. The symmetric argument gives $F_2\succeq_{\rm NE}F_1$.
\end{proof}

\subsection{Proof of Theorem~\ref{Theorem: stochastic loop cover dominance}}

\begin{proof} 
For necessity, local refinement in every CKC follows from Lemma~\ref{Lemma - single CKC refinement}. The existence of an $F_2$-cover for every irreducible $F_1$-loop follows from Lemma~\ref{Lemma - cover necessity}. If such a loop has a unique cover, order preservation follows from Lemma~\ref{Lemma - unique cover order preserving}.

We prove sufficiency. Fix an arbitrary $F_2$-measurable experiment $\tau_2$ on a finite signal set $S_2$. Let $\mathcal C$ be the CKC partition. We first construct an $F_1$-measurable experiment $\tau_1$ that induces the same distribution of posteriors for a single decision maker whose information partition is $\mathcal C$.
For each CKC $C$ and each $F_1$-atom $A$ with $A\cap C\neq\emptyset$, define $p_{A,C}\in\Delta(S_2)$ by $p_{A,C}:=\tau_2(\cdot|\omega)$ for any $\omega\in A\cap C$. This is well-defined because $F_1$ refines $F_2$ inside $C$.

Construct the bipartite incidence graph $\Gamma$: its vertices are the CKCs and the $F_1$-atoms, and $A$ is connected to $C$ iff $A\cap C\neq\emptyset$. By the separated-loop assumption, no vertex of $\Gamma$ lies on two distinct simple cycles. Hence Lemma~\ref{Lemma - separated incidence graph} below applies to each connected component of $\Gamma$.

We assign to each $F_1$-atom $A$ a law $q_A$ such that, for every CKC $C$, there is a law
$\nu_C$ with $q_A=p_{A,C}\otimes\nu_C$ for every atom $A$ adjacent to $C$. This condition means that, inside $C$, the first coordinate reproduces the experiment $\tau_2$, while the remaining coordinates are common noise.

Fix one connected component of $\Gamma$. First suppose that it is a tree. Denote its CKCs by $T$. Use one coordinate \(S_D=S_2\) for each CKC \(D \in T\), with signals $s_D \in S_D$. For an \(F_1\)-atom \(A\) and a CKC \(D\in T\), let \(\phi_D(A)\) be the unique \(F_1\)-atom adjacent to \(D\) that lies on the path from \(D\) to \(A\). Define
\[
        q_A((s_D)_D)=\prod_{D\in T} p_{\phi_D(A),D}(s_D).
\]
Fix a CKC \(C\). If \(A\) is adjacent to \(C\), then the \(C\)-coordinate of \(q_A\) is
\(p_{A,C}\). For every \(D\neq C\), all \(F_1\)-atoms adjacent to \(C\) remain connected to \(C\) after deleting \(D\). Hence, the unique path from \(D\) to any atom adjacent to \(C\) leaves \(D\) through the same neighboring \(F_1\)-atom. Therefore \(\phi_D(A)\) is the same for all \(F_1\)-atoms \(A\) adjacent to \(C\). Thus all coordinates other than \(C\) are common across atoms adjacent to \(C\), and the required property holds on tree components.

Now suppose that the component contains at least one simple cycle. Let $\mathcal B$ be the set of simple cycles in this component.  For each $B\in\mathcal B$, write $B=A_0,C_1,A_1,\ldots,A_{m-1},C_m,A_0$, and put $A_m=A_0$. For every $i$, choose $\omega_i\in A_{i-1}\cap C_i$ and $\bar\omega_i\in A_i\cap C_i$. Since $B$ is simple, these $2m$ states are distinct and $(\omega_1,\bar\omega_1,\ldots,\omega_m,\bar\omega_m)$ is an irreducible $F_1$-loop whose connecting atoms are $A_0,\ldots,A_{m-1}$. By assumption, this loop has an order-preserving $F_2$-cover. Applying Lemma~\ref{Lemma - single loop construction} to this cycle and $\tau_2$ gives a signal set $S_B$ and laws $h_A^B$ for the cycle atoms such that, for each cycle CKC $C_i$, there is a law $\eta_i^B$ with $h_A^B=p_{A,C_i}\otimes\eta_i^B$ for the two cycle atoms $A$ adjacent to $C_i$.

For any off-cycle atom $A$ adjacent to a cycle CKC $C_i$ of $B$, set $h_A^B:=p_{A,C_i}\otimes\eta_i^B$. This is unambiguous by Lemma~\ref{Lemma - separated incidence graph}(i).

Let $\mathcal R_B$ be the set of $F_1$-atoms for which $h_A^B$ has just been defined; that is, the cycle atoms of $B$ and the off-cycle atoms adjacent to CKCs of $B$. For any atom $A$ in the component, let $R_B(A)$ be the unique element of $\mathcal R_B$ in the component containing $A$ after deleting the cycle CKC vertices of $B$ and their incident edges. Existence and uniqueness of this root follow from Lemma~\ref{Lemma - separated incidence graph}(ii).

Let $T$ now denote the set of CKCs in the component that do not lie on any simple cycle. For each $D\in T$, use one coordinate \(S_D=S_2\). For an atom \(A\) and \(D\in T\), let
\(\phi_D(A)\) be the unique atom adjacent to \(D\) that lies in the component of
\(\Gamma\setminus\{D\}\) containing \(A\). This is well-defined by Lemma~\ref{Lemma - separated incidence graph}(iii).

For the whole component, use the signal space $\left(\prod_{B\in\mathcal B} S_B\right)\times \left(\prod_{D\in T} S_D\right)$, and define, for every atom \(A\) in the component,
\[
    q_A = \left(\bigotimes_{B\in\mathcal B} h_{R_B(A)}^B\right) \otimes \left(\bigotimes_{D\in T} p_{\phi_D(A),D}\right).
\]
Since $\mathcal B$ and $T$ are finite and every tensor factor is a probability law, $q_A$ is a well-defined probability law.

Fix a CKC $C$ and two atoms $A,A'$ adjacent to $C$. Suppose first that $C$ lies on a cycle $B_C$. Since $A,A'\in\mathcal R_{B_C}$, we have $R_{B_C}(A)=A$ and $R_{B_C}(A')=A'$. Hence the $B_C$-coordinate gives $p_{A,C}\otimes\eta_C^{B_C}$ at $A$ and $p_{A',C}\otimes\eta_C^{B_C}$ at $A'$.

For every other cycle $B\neq B_C$, the path $A-C-A'$ remains after deleting the CKC vertices of $B$, because distinct cycles are vertex-disjoint. Hence $R_B(A)=R_B(A')$. Likewise, for every $D\in T$, the path $A-C-A'$ remains in $\Gamma\setminus\{D\}$, so $\phi_D(A)=\phi_D(A')$. Thus every coordinate other than the $B_C$-coordinate is common noise inside $C$.

Now suppose that $C\in T$. Then $\phi_C(A)=A$ and $\phi_C(A')=A'$, so the $C$-coordinate gives $p_{A,C}$ and $p_{A',C}$. For every cycle $B$, the path $A-C-A'$ remains after deleting the CKC vertices of $B$, since $C$ lies on no cycle. Similarly, for every $D\in T\setminus\{C\}$, the path remains after deleting $D$. Therefore $R_B(A)=R_B(A')$ and $\phi_D(A)=\phi_D(A')$ for every non-target coordinate. Consequently, for every CKC $C$ there exists a law $\nu_C$ such that $q_A= p_{A,C} \otimes \nu_C$ for every atom $A$ adjacent to $C$.

Doing this component by component and taking the product across components gives one finite signal space and one law $q_A$ for every relevant $F_1$-atom $A$. Define $\tau_1(\cdot|\omega):= q_{F_1(\omega)}$. Then $\tau_1$ is $F_1$-measurable. Moreover, for every CKC $C$, there exists a common-noise law $\nu_C$ such that $q_A=p_{A,C}\otimes\nu_C$ for every $F_1$-atom $A$ intersecting $C$. Therefore, in the single-DM environment with information partition $\mathcal C$, $\tau_1$ and $\tau_2$ induce the same distribution of posteriors.

Finally, return to the original multi-agent environment. The construction preserves the joint law of \((\omega,\mu(\cdot\mid C(\omega),\tau,s))\), since within each CKC the experiment \(\tau_1\) reproduces \(\tau_2\) up to common noise. Given the realized state and the CKC-level posterior, each player's posterior is uniquely determined by her information cell \(\Pi_i(\omega)\). Hence, \(\tau_1\) reproduces \(\tau_2\) within each CKC up to common noise. Hence \(\overline{\mu}_{\tau_1\mid C}=\overline{\mu}_{\tau_2\mid C}\) for every CKC \(C\). By Proposition~\ref{Prop - Charact in terms of signaling funct}, \(F_1\succeq_{\rm NE}F_2\).
\end{proof}

\begin{lemma}[Separated incidence graph] \label{Lemma - separated incidence graph}
Let $\Gamma$ be a connected bipartite graph whose vertices are CKCs and atoms, and suppose that no vertex belongs to two distinct simple cycles. Fix a simple cycle $B$, let $\mathcal C(B)$ be its CKC
vertices, and let $\mathcal R_B$ be the set of atom vertices adjacent
to some vertex of $\mathcal C(B)$.

\begin{enumerate}[(i)]
    \item Every atom outside $B$ is adjacent to at most one CKC vertex of $B$.
    \item Every component of $\Gamma\setminus\mathcal C(B)$ contains exactly one element of $\mathcal R_B$.
    \item If a CKC vertex $D$ lies on no simple cycle, then every component of $\Gamma\setminus\{D\}$ contains exactly one atom adjacent to $D$.
\end{enumerate}
\end{lemma}

\begin{proof}
For (i), if an atom outside $B$ were adjacent to two CKC vertices of $B$, these two edges together with an arc of $B$ would form another simple cycle sharing vertices with $B$.

For (ii), connectedness implies that every component of $\Gamma\setminus\mathcal C(B)$ contains at least one element of $\mathcal R_B$. If one component contained two roots, choose two at minimum graph distance and let $P$ be a shortest path between them, so that the interior of $P$ contains no root. Each endpoint either lies on $B$ or is joined to a CKC vertex of $B$, uniquely in the latter case by part~(i). Adding these attachment edges and, when the attachment points are distinct, an arc of $B$ produces a simple cycle distinct from $B$ that shares a vertex with $B$, a contradiction.

For (iii), connectedness implies that every component of $\Gamma\setminus\{D\}$ contains at least one neighbor of $D$. If one component contained two such neighbors, a path between them together with their two incident edges to $D$ would form a simple cycle through $D$, a contradiction.
\end{proof}

\subsection{Proof of Corollary~\ref{Theorem: NI leads to dominance}}

\begin{proof}
Fix an \(F_2\)-measurable experiment \(\tau_2\), and let \(\Gamma\) be the bipartite incidence graph between the CKCs and the \(F_1\)-atoms. It suffices to work on one connected component \(H\) of \(\Gamma\), since the resulting experiments can be combined across components by the same product construction used in previous proofs.

For every incidence \(A-C\), define \(p_{A,C}:=\tau_2(\cdot\mid\omega)\) where $\omega\in A\cap C$. This is well-defined because \(F_1\) refines \(F_2\) inside \(C\). For each CKC \(D\) in \(H\) and each component \(K\) of \(H\setminus\{D\}\), choose an \(F_1\)-atom \(A_D(K)\) adjacent
to \(D\) and contained in \(K\), and set \( r_{D,K}:=p_{A_D(K),D}\). This is independent of the choice of \(A_D(K)\). Indeed, if two atoms \(A,A'\) adjacent to \(D\) belong to the same component of \(H\setminus\{D\}\), a shortest path between them, closed through \(D\), induces an \(F_1\)-loop. Its \(D\)-pair is \(F_2\)-non-informative, so \(p_{A,D}=p_{A',D}\).

For every \(F_1\)-atom \(A\) in \(H\), let \(K_D(A)\) denote its component in \(H\setminus\{D\}\), and define \(q_A:=\bigotimes_{D\in\mathcal C(H)}r_{D,K_D(A)}\). If \(A\) meets \(C\), then its \(C\)-coordinate is \(p_{A,C}\). Moreover, if \(A,A'\) both meet \(C\), then for every \(D\neq C\)
the path \(A-C-A'\) survives deletion of \(D\), so \(K_D(A)=K_D(A')\). Hence there is a law \(\nu_C\) such that \( q_A=p_{A,C}\otimes\nu_C\) for every \(A\) meeting \(C\).

Set \( \tau_1(\cdot\mid\omega):=q_{F_1(\omega)}\) and repeat the construction on the other components of \(\Gamma\). The resulting experiment is \(F_1\)-measurable and reproduces \(\tau_2\) within every CKC up to common noise. Proposition~\ref{Prop - Charact in terms of signaling funct} therefore implies \(F_1\succeq_{\rm NE}F_2\).
\end{proof}

\subsection{Proof of Proposition~\ref{Proposition: two CKCs}}

\begin{proof}
Necessity of local refinement follows from Lemma~\ref{Lemma - single CKC refinement}. Necessity of the irreducible cover condition follows from Lemma~\ref{Lemma - cover necessity}. We show that, with two CKCs, this implies that every \(F_1\)-loop is \(F_2\)-covered. 

Let the two CKCs be \(C_1\) and \(C_2\) and denote \(\omega_1,\omega_1'\in C_1\) and \(\omega_2,\omega_2'\in C_2\). Call an \(F_1\)-atom cross-CKC if it intersects both \(C_1\) and \(C_2\). By local refinement, if \(F_1(\omega_1)=F_1(\omega_2)\), then \(F_1(\omega_1)\cap C_1\subseteq F_2(\omega_1)\) and \(F_1(\omega_1)\cap C_2\subseteq F_2(\omega_2)\). 

The case with at most one cross-CKC $F_1$-atom is immediate, so take two distinct cross-CKC $F_1$-atoms, \(F_1(\omega_1)=F_1(\omega_2)\) and \(F_1(\omega_1')=F_1(\omega_2')\). These two atoms form an irreducible \(F_1\)-loop, so are \(F_2\)-covered. This produces the following dichotomy: either the two pairs themselves form an \(F_2\)-loop or both CKC-pairs are \(F_2\)-non-informative. This holds for \emph{any} two cross-CKC $F_1$-atoms, so either all cross-CKC \(F_1\)-atoms as above are contained in cross-CKC \(F_2\)-atoms, or all cross-CKC \(F_1\)-atoms are contained in the same $F_2$ atom within every CKC (namely, \(F_2(\omega_1)=F_2(\omega_1')\) and \(F_2(\omega_2)=F_2(\omega_2')\)). Indeed, consider the two cross-CKC $F_1$-atoms above along with a third cross-CKC $F_1$-atom. These atoms generate 3 irreducible $F_1$-loops. If one of these atoms is contained in a cross-CKC \(F_2\)-atom, then the cover condition implies that the other two are also contained in \(F_2\)-atoms. 
So either every \(F_1\)-loop is itself an \(F_2\)-loop, or every CKC-pair in any \(F_1\)-loop is \(F_2\)-non-informative, so the loop is covered by putting all pairs in \(J\). Therefore every \(F_1\)-loop is \(F_2\)-covered.

For sufficiency, assume local refinement and that every \(F_1\)-loop is \(F_2\)-covered. If there is at most one cross-CKC $F_1$-atom, every $F_1$-loop is $F_1$-non-informative and hence $F_2$-non-informative by local refinement. Corollary~\ref{Theorem: NI leads to dominance} therefore implies $F_1\succeq_{\rm NE}F_2$. Otherwise, the same dichotomy applies. In the first case, \(F_1\) refines \(F_2\) globally, so every \(F_2\)-measurable experiment is already \(F_1\)-measurable, and \(F_1\succeq_{\rm NE}F_2\). In the second case, every \(F_1\)-loop is \(F_2\)-non-informative, so Corollary~\ref{Theorem: NI leads to dominance} implies \(F_1\succeq_{\rm NE}F_2\).
\end{proof}

\newpage

%%%%%%%%%%%%%%%%%%%%%%%%%%%%%%%%%%%%%%%%%%%%%%%%%%%%%%

\section{Online appendix}

\subsection{Finer in every CKC can be inferior: a fact-checking example} \label{Appendix: polarization example}

In a political debate watched by two citizens, a fair coin toss decides who gets to make the final claim in the debate: either the incumbent makes a budget (B) claim, or the challenger makes a security (S) claim. Any claim is rated True (T), Mostly true (M), or False (F). If a claim made by an individual is rated T or M, it favors that individual, but if it is rated F, it favors the opponent. The context itself is common knowledge, so the two CKCs are \(C_B=\{\omega_1,\ldots,\omega_6\}\) and \(C_S=\{\omega_7,\ldots,\omega_{12}\}\). The political impact is shown by the fill color in Figure~\ref{fig:factchecker-grid}: blue means \(+\), and red means \(-\). Rows are CKCs and columns are verdict labels.

The network in charge of this debate can employ two fact-checking report formats: an issue-specific impact format and a verdict format. The issue-specific impact format \(\mathcal F^{\mathrm{imp}}\) reports whose side is favored by the facts within the public context. Its atoms are \(B^+=\{\omega_1,\omega_2,\omega_3,\omega_4\}\), \(B^-=\{\omega_5,\omega_6\}\), \(S^-=\{\omega_7,\omega_8,\omega_9,\omega_{10}\}\), and \(S^+=\{\omega_{11},\omega_{12}\}\). This format has no atom crossing CKCs. The verdict format \(\mathcal F^{\mathrm{ver}}\) reports the fact-check verdict. Its atoms are \(V_T=\{\omega_1,\omega_2,\omega_7,\omega_8\}\), \(V_M=\{\omega_3,\omega_4,\omega_9,\omega_{10}\}\), and \(V_F=\{\omega_5,\omega_6, \omega_{11}, \omega_{12}\}\).  The context is common knowledge, including to the network, but each report format is modeled as a binding reporting technology. Under the verdict format, the experiment must remain \(\mathcal F^{\mathrm{ver}}\)-measurable and cannot additionally condition its report law on the context.

Notably, the verdict format is \emph{strictly finer} than the issue-specific impact format inside each CKC, row by row. But every verdict column flips political impact across rows: True and Mostly true are incumbent-favorable in Budget but challenger-favorable in Security, while False is challenger-favorable in Budget but incumbent-favorable in Security.

\begin{figure}[ht]
\centering
\renewcommand{\twostatecell}[5]{%
  \begin{minipage}{2.40cm}
  \(\boldsymbol{#1}\)\\[5pt]
  \(#2\)\hfill \(#3\)\\[4pt]
  \(#4\)\hfill \(#5\)
  \end{minipage}%
}
\begin{tikzpicture}[
  every node/.style={font=\small},
  pluscell/.append style={minimum width=3.10cm},
  minuscell/.append style={minimum width=3.10cm}
]  % Coordinates
  % Coordinates
  \def\xT{0}
  \def\xM{3.50}
  \def\xF{7.00}
  \def\yB{0}
  \def\yS{-3.05}

  % Row labels
  \node[font=\bfseries\large, anchor=east, align=right]
    at (-1.95,\yB)
    {Budget claim\\[-2pt]by incumbent};

  \node[font=\bfseries\large, anchor=east, align=right]
    at (-1.95,\yS)
    {Security claim\\[-2pt]by challenger};

  % Budget row
  \node[pluscell]  (BT) at (\xT,\yB)
  {\twostatecell{+}{\omega_1}{1}{\omega_2}{5}};
  \node[pluscell]  (BM) at (\xM,\yB)
  {\twostatecell{+}{\omega_3}{1}{\omega_4}{5}};
  \node[minuscell] (BF) at (\xF,\yB)
  {\twostatecell{-}{\omega_5}{8}{\omega_6}{12}};

  % Security row
  \node[minuscell] (ST) at (\xT,\yS)
  {\twostatecell{-}{\omega_7}{4}{\omega_8}{6}};
  \node[minuscell] (SM) at (\xM,\yS)
  {\twostatecell{-}{\omega_9}{4}{\omega_{10}}{6}};
  \node[pluscell]  (SF) at (\xF,\yS)
  {\twostatecell{+}{\omega_{11}}{2}{\omega_{12}}{10}};

  % Column labels, anchored above the top-row cells
  \node[font=\bfseries\large, align=center, anchor=south, inner sep=1pt]
    at ([yshift=12pt]BT.north) {True\\[-2pt]\(\boldsymbol{T}\)};
  \node[font=\bfseries\large, align=center, anchor=south, inner sep=1pt]
    at ([yshift=12pt]BM.north) {Mostly true\\[-2pt]\(\boldsymbol{M}\)};
  \node[font=\bfseries\large, align=center, anchor=south, inner sep=1pt]
    at ([yshift=12pt]BF.north) {False\\[-2pt]\(\boldsymbol{F}\)};

\end{tikzpicture}
\caption{\footnotesize Rows are common-knowledge components, columns are fact-check verdict labels, and color is political impact: blue \(+\), incumbent-favorable; red \(-\), challenger-favorable. Numbers \(=\) multiplicities, normalized by \(32\) within each context. The verdict format refines political impact within each row, but every verdict column flips political impact across rows.}
\label{fig:factchecker-grid}
\end{figure}

Formally, each common-knowledge component has prior probability \(1/2\). Conditional probabilities within each component are obtained by normalizing the multiplicities in Figure~\ref{fig:factchecker-grid} by \(32\). The citizens' partitions are
\[
\begin{aligned}
\Pi_1 &= \bigl\{
  \{\omega_1,\omega_2,\omega_3,\omega_4,\omega_6\},
  \{\omega_5\},
  \{\omega_8,\omega_{10},\omega_{11},\omega_{12}\},
  \{\omega_7,\omega_9\}\bigr\},\\
\Pi_2 &= \bigl\{
  \{\omega_1,\omega_3\},
  \{\omega_2,\omega_4,\omega_5,\omega_6\},
  \{\omega_{11}\},
  \{\omega_7,\omega_8,\omega_9,\omega_{10},\omega_{12}\}
  \bigr\}.
\end{aligned}
\]
These partitions generate exactly \(C_B\) and \(C_S\).

Let \(D(\tau)=\mathbb E_\mu[\|\mu^1_{\tau}-\mu^2_{\tau}\|_1]\) be the \(\ell_1\)-expected distance between the citizens' posteriors. With the weights above, the impact format can use the binary experiment
\[
\tau^{\mathrm{imp}}(s_1\mid \omega)=
\begin{cases}
0, & \text{if } \omega\in B^+\cup S^+,\\[2pt]
\tfrac{3}{4}, & \text{if } \omega\in B^-\cup S^-,
\end{cases}
\qquad
\tau^{\mathrm{imp}}(s_2\mid \omega)=1-\tau^{\mathrm{imp}}(s_1\mid \omega).
\]
This experiment obtains \(D(\tau^{\mathrm{imp}})=31/40\). By contrast, every verdict-measurable experiment satisfies \(D(\tau)\ge 2879/3600\), and one can show that this bound is tight and attained by the binary experiment
\[
\tau^{\mathrm{ver}}(s_1\mid \omega)=
\begin{cases}
\tfrac{1}{5}, & \text{if } \omega\in\ V_T\cup V_M ,\\[2pt]
\tfrac{4}{5}, & \text{if } \omega\in V_F,
\end{cases}
\qquad
\tau^{\mathrm{ver}}(s_2\mid \omega)=1-\tau^{\mathrm{ver}}(s_1\mid \omega).
\]

Therefore, a fact-checking format can be strictly finer inside every CKC and still be strictly worse for depolarization. The coarser impact format is disconnected across CKCs; the obstruction comes entirely from the globally glued verdict labels, which create the loop. Moreover, garbling can depolarize strictly more than full revelation within each CKC.

\paragraph{Verification.}
The multiplicities are chosen so that both issues induce the same four-state depolarization block after marginalizing the split verdict states. This marginalization is applied to the citizens' posterior beliefs and does not require the verdict format to identify the aggregated states. In Budget, aggregate \(\omega_1,\omega_3\) into base state \(1\), keep \(\omega_5\) as base state \(2\), aggregate \(\omega_2,\omega_4\) into base state \(3\), and keep \(\omega_6\) as base state \(4\). In Security, keep \(\omega_{11}\) as base state \(1\), aggregate \(\omega_7,\omega_9\) into base state \(2\), keep \(\omega_{12}\) as base state \(3\), and aggregate \(\omega_8,\omega_{10}\) into base state \(4\). In both components, the induced prior is \((1/16,4/16,5/16,6/16)\), Citizen 1's induced partition is \(\{\{1,3,4\},\{2\}\}\), and Citizen 2's induced partition is \(\{\{1\},\{2,3,4\}\}\).

Under the issue-specific impact format, \(B^+\) and \(S^+\) are \(\{1,3\}\) in their respective components, while \(B^-\) and \(S^-\) are \(\{2,4\}\). Hence, the binary experiment
\(\tau^{\mathrm{imp}}(s_1\mid B^+) =\tau^{\mathrm{imp}}(s_1\mid S^+)=0\) and
\(\tau^{\mathrm{imp}}(s_1\mid B^-) =\tau^{\mathrm{imp}}(s_1\mid S^-)=3/4\), with \(s_2\) complementary, is the aligned four-state experiment in both components. Because it assigns the same likelihood to the states within every aggregated pair, marginalization is lossless, and direct
calculation gives \(D(\tau^{\mathrm{imp}})=31/40\).

Now take any verdict-measurable experiment. Write \(p_T(s)=\tau(s\mid V_T)\), \(p_M(s)=\tau(s\mid V_M)\), \(q_F(s)=\tau(s\mid V_F)\), and \(\bar p(s)=(p_T(s)+p_M(s))/2\). The induced marginal experiment in Budget uses likelihood \(\bar p(s)\) on \(\{1,3\}\) and \(q_F(s)\) on \(\{2,4\}\), while the induced marginal experiment in Security uses likelihood \(q_F(s)\) on \(\{1,3\}\) and \(\bar p(s)\) on \(\{2,4\}\). Thus every verdict-measurable experiment induces the looped four-state experiment with reversed likelihoods across the two components.

To calculate its minimum, consider one signal with likelihood \(a\) on \(\{1,3\}\) and \(b\) on \(\{2,4\}\).  For \(a+b>0\), the citizens' posteriors on their non-singleton cells are
\[
u(a,b)=\frac{(a,0,5a,6b)}{6a+6b},
\qquad
v(a,b)=\frac{(0,4b,5a,6b)}{5a+10b}.
\]
The signal's contribution to disagreement in one component is
\[
\begin{aligned}
d(a,b):={}&\frac{a}{16}\|u(a,b)-e_1\|_1
          +\frac{4b}{16}\|e_2-v(a,b)\|_1\\
         &+\frac{5a+6b}{16}\|u(a,b)-v(a,b)\|_1,
\end{aligned}
\]
with \(d(0,0)=0\). Since the two components reverse the likelihoods and are equally likely, put \(G(a,b):=[d(a,b)+d(b,a)]/2\). Therefore, with \(a_s=\bar p(s)\) and \(b_s=q_F(s)\), marginalization implies \(D(\tau)\geq\sum_sG(a_s,b_s)\).

Let \(c=2879/7200\). Since \(G\) is symmetric and homogeneous, for \(a+b>0\) set \(x=a/(a+b)\) and assume \(x\leq1/2\). If \(0\leq x\leq1/5\), direct simplification gives
\(\frac{G(a,b)}{a+b}-c =\frac{(1-5x)(62+761x-66x^2-45x^3)}{7200(2-x)(1+x)}\geq0\). If \(1/5\leq x\leq1/2\), it gives \(\frac{G(a,b)}{a+b}-c =\frac{(5x-1)(4-5x)(86+3x-3x^2)}{3600(2-x)(1+x)}\geq0\). Hence \(G(a,b)\geq c(a+b)\) for all \(a,b\geq0\). Since
\(\sum_sa_s=\sum_sb_s=1\), it follows that \(D(\tau)\geq c\sum_s(a_s+b_s)=2879/3600\).

For the proposed binary verdict experiment, the likelihood pairs are
\((1/5,4/5)\) and \((4/5,1/5)\), so equality holds in the pointwise
bound. Moreover, \(p_T=p_M\), making the marginalization lossless.
Hence
\(\inf_{\tau:\,\mathcal F^{\mathrm{ver}}\text{-measurable}}D(\tau)
=2879/3600>31/40\).

\subsection{A takeover bidding war}\label{Appendix: takeover bidding war}

Consider the information structure given in the mediation-game example (Section~\ref{Section - A simple example}), where the two players are firms competing in a takeover contest for a target firm. We interpret the state space as representing the possible synergies from the acquisition: Firm~$1$ has higher synergies in $\o_1$ and $\o_3$, while Firm~$2$ has higher synergies in $\o_2$ and $\o_4$. Each firm has two actions, $R$ and $W$, where $R$ denotes raising its offer and $W$ denotes withdrawing. To simplify the exposition, normalize the payoff from withdrawal to $0$. A high-synergy bidder obtains $1$ from raising its offer when its rival withdraws and $1/2$ when both raise; a low-synergy bidder obtains $-1$ and $-2$, respectively. Thus, an informed high-synergy bidder chooses $R$, whereas an informed low-synergy bidder chooses $W$.

Denote by $p$ the probability that an uninformed firm assigns to having high synergies. It is easy to verify that it raises its offer iff $p\geq 2/3$. Thus, neither a null report nor a fully revealing report generates a bidding war. Under a null report, the uncertain bidder withdraws, while under a fully revealing report, the low-synergy bidder withdraws and the high-synergy bidder raises.

The two analysts have partitions
\[
F_1=\bigl\{\{\omega_1,\omega_3\},
          \{\omega_2,\omega_4\}\bigr\},
\qquad
F_2=\bigl\{\{\omega_1,\omega_4\},
          \{\omega_2,\omega_3\}\bigr\}.
\] 
Again, both analysts are perfectly informed within every CKC, but connect the two CKCs differently. Following the same reasoning as in Section~\ref{Section - A simple example}, Analyst~2 can send a partially informative signal $s$ with probability $1$ in $\o_1$ and $\o_4$ and probability $1/2$ in $\o_2$ and $\o_3$. This experiment is $F_2$-measurable, and conditional on $s$, the uncertain bidder assigns probability $2/3$ to having high synergies in either CKC. Therefore, there exists an equilibrium in which it raises its offer. In $\o_2$ and $\o_3$, that bidder actually has low synergies, while its informed high-synergy rival also raises. Thus, the ex ante probability of a bidding war is $1/4$. The same analysis as in the mediation-game example shows that the other analyst can induce a bidding war with an ex ante probability of at most $1/6$.
 
We can thus show that selective disclosure can generate a bidding war that arises under neither silence nor full transparency, and that two locally equivalent analysts can differ solely because their information connects the CKCs differently.

\subsection{Tree mimicry} \label{Appendix: tree mimicry}

The following result provides an edge-based independent-coordinate construction. Each edge of a tree is assigned a binary experiment, possibly on a different signal space. These experiments can be represented simultaneously by laws on a common signal space, with the additional coordinates acting as common noise.

\begin{lemma}[Tree mimicry] \label{Lemma - tree mimicry}
Let \(T=(V,E)\) be a finite directed tree. 
For each edge \(e=(u,v)\in E\), fix a finite signal set \(S_e\) and two laws \(\alpha_e,\beta_e\in\Delta(S_e)\). 
Then there exist a finite signal set \(S_T\) and laws \(q_v\in\Delta(S_T)\), \(v\in V\), such that for every edge \(e=(u,v)\), the ordered pair \((q_u,q_v)\) is a posterior-equivalent extension of \((\alpha_e,\beta_e)\).
\end{lemma}

\begin{proof}
Let \(S_T=\prod_{e\in E}S_e\). For an edge \(e=(u,v)\), deleting \(e\) leaves two connected components, so denote by \(T_e^-\) the component containing \(u\) and by \(T_e^+\) the component containing \(v\). 
For \(x\in V\), define \(q_x\in\Delta(S_T)\) by
\[
q_x((s_f)_{f\in E})=
\prod_{e\in E}
\begin{cases}
\alpha_e(s_e),& x\in T_e^-,\\
\beta_e(s_e),& x\in T_e^+ .
\end{cases}
\]
This is a probability law because it is a product of probability laws. 
Fix \(e=(u,v)\). 
If \(f\neq e\), then \(u\) and \(v\) lie in the same component of \(T\setminus\{f\}\), so the \(f\)-coordinate has the same law under \(q_u\) and \(q_v\). 
On coordinate \(e\), the law is \(\alpha_e\) under \(q_u\) and \(\beta_e\) under \(q_v\). Hence \((q_u,q_v)\) is \((\alpha_e,\beta_e)\) multiplied by the same law on the coordinates \(E\setminus\{e\}\), and is therefore a posterior-equivalent extension.
\end{proof}

\subsection{More than one CKC}\label{Subsection_more than one CKC two examples}

As the benchmark Corollary~\ref{Corollary: stochastic benchmark cases} states, under a unique CKC, the partition-refinement condition ensures that Oracle $1$ can produce the \emph{exact} same experiment as Oracle $2$. This conclusion, however, hinges on the existence of a unique CKC. In case there are several CKCs, Oracle $1$ may need to follow a different experiment in order to match the distribution on posteriors generated by $\tau_2$. Namely, $\tau_1$ may require more signals than $\tau_2$, even if both oracles have the same (complete) information in every CKC. Let us provide a concrete example for this.

\begin{example} A mimicking experiment $\tau_1$ may require more signals than $\tau_2$.
\end{example}

Consider a uniformly distributed state space $\Omega =\{\o_1,\o_2,\o_3,\o_4\}$, with two players whose private information is $\Pi_1=\{\{\o_1,\o_2\},\{\o_3\},\{\o_4\}\}$ and $\Pi_2=\{\{\o_1\},\{\o_2\},\{\o_3,\o_4\}\}$.
The oracles have the following partitions $F_1 = \{\{\o_1,\o_3\}, \{\o_2\},\{\o_4\}\}$ and $F_2 = \{\{\o_1\},\{\o_3\}, \{\o_2,\o_4\}\}$.
Notice that there are two CKCs, $\{\o_1,\o_2\}$ and $\{\o_3,\o_4\}$, and both oracles have complete information in each of these components.
That is, $F_1$ refines $F_2$ in every CKC, and vice versa.

Consider the experiment $\tau_2$ given in Figure \ref{fig:Table of tau2 in example 2}.
Notice it is $F_2$-measurable, as $\tau_2(s|\o_2)=\tau_2(s|\o_4)$ for every signal $s$, but not $F_1$-measurable.
\begin{figure}[th!]
\centering

\medskip

\begin{tabular}{c|c|c|c|}
    $\tau_2(s|\o)$ & $s_1$ & $s_2$ & $s_3$ \\
\hline
$\o_1$ & 0 & 1/2 & 1/2 \\
\hline
$\o_2$ & 1/3 & 2/3 & 0 \\
\hline
$\o_3$ & 0 & 2/3 & 1/3 \\
\hline
$\o_4$ & 1/3 & 2/3 & 0 \\
\hline
\end{tabular}
\caption{ \footnotesize A stochastic $F_2$-measurable experiment of Oracle $2$.}
\label{fig:Table of tau2 in example 2}
\end{figure}

The set of posterior profiles generated by \(\tau_2\) is
\[
{\rm Post}(\tau_2) =
\left\{\!\begin{aligned}
&(e_i, e_i), && \forall \ 1 \leq i \leq 4, \\[1ex]
&\left(\left(\tfrac{3}{7}, \tfrac{4}{7}, 0, 0\right), e_j\right), && j = 1, 2, \\[1ex]
&\left(e_k, (0, 0, \tfrac{1}{2}, \tfrac{1}{2})\right), && k = 3, 4
\end{aligned}\right\},
\]
where the \(i^{\rm th}\) coordinate of \(e_i\) is \(1\). The set \({\rm Post}(\tau_2)\) records the posterior types that can arise, but mimicry also requires matching their conditional probabilities
within each CKC. Under \(\tau_2\), in \(C_{12}=\{\omega_1,\omega_2\}\), the posterior types
\(e_1\), \((3/7,4/7,0,0)\), and \(e_2\) occur with probabilities \(1/4\), \(7/12\), and \(1/6\), respectively. In \(C_{34}=\{\omega_3,\omega_4\}\), the posterior types \(e_3\), \((0,0,1/2,1/2)\), and \(e_4\) occur with probabilities \(1/6\), \(2/3\), and \(1/6\), respectively.

With only three signals, each posterior type in either CKC must be generated by exactly one signal, so the common signal labels define a bijection between the three types in the two CKCs. Under the
uniform prior, a signal that occurs with conditional probability \(m\) in a two-state CKC and induces posterior probability \(q\) on its first state has likelihood \(2mq\) at that state. Therefore, the three signals must have likelihoods \(\{1/2,1/2,0\}\) at \(\omega_1\), but
\(\{2/3,1/3,0\}\) at \(\omega_3\). Since \(F_1(\omega_1)=F_1(\omega_3)\), \(F_1\)-measurability requires these likelihoods to coincide signal by signal. No bijection between the two multisets satisfies this requirement. Hence no three-signal \(F_1\)-measurable experiment can reproduce the required distributions of CKC posteriors.

Oracle~1 therefore requires a fourth signal, as in Figure~\ref{fig:Table of tau1 in example 2}. Thus, even when the oracles' partitions refine one another in every CKC, a mimicking
experiment may require more signals than the experiment it mimics.

\begin{figure}[th!]
\centering
\begin{tabular}{c|c|c|c|c|}
    $\tau_1(s|\o)$ & $s_3$ & $s_4$ & $s_5$ & $s_6$ \\
\hline
$\o_1$ & $1/2$ & 1/3 & $0$   & $1/6$   \\
\hline
$\o_2$ & $2/3$  & $0$ & $1/3$ & $0$     \\
\hline
$\o_3$ & $1/2$  & $1/3$ &  $0$ & $1/6$  \\
\hline
$\o_4$ & $1/2$ & $0$ & $1/3$ & $1/6$   \\
\hline
\end{tabular}
\caption{ \footnotesize An experiment $\tau_1$ with $4$ signals.}\label{fig:Table of tau1 in example 2}
\end{figure}

\begin{remark}
    Note that dominance does not imply refinement in general. To see this, consider the information structure $ \Pi_1 = \Pi_2 = \{\{\o_1,\o_2\},\{\o_3,\o_4\}\}, F_1 = \{\{\o_1,\o_2,\o_3\},\{\o_4\}\}$ and $F_2  = \{\{\o_1,\o_2\},\{\o_3\},\{\o_4\}\}$. For every experiment $\tau_2$, one can devise an experiment $\tau_1$ that yields the same distribution over posterior beliefs. Evidently, Oracle $2$ provides the players with no additional information regarding states $\o_1$ and $\o_2$, and this allows Oracle $1$ to replicate $\tau_2$ on $\o_3$ and $\o_4$, accordingly.
\end{remark}

\subsection{Balancedness and the likelihood-ratio obstruction} \label{Appendix: balanced-cover motivation}

This subsection provides additional motivation for the cover criterion in Appendix~\ref{Appendix: loop algebra}. Consider a non-balanced $F_1$-loop as in Figure~\ref{fig: F1 loop and non-balanced loop}(b), and suppose that $A$ and $B$ are $F_2$-measurable sets with $\#(A\to B)\neq \#(B\to A)$. For example, if $A=\{\omega_1,\omega_2,\omega_3\}$ and $B=\{\overline\omega_1, \overline\omega_2, \overline\omega_3\}$, then $\#(A\to B)=3$ and $\#(B\to A)=0$. Let $s$ be a signal and define an $F_2$-measurable experiment by
\[
    \tau_2(s\mid\omega)=\frac12-\frac14\mathbf 1_{\{\omega\in A\}}.
\]
Then each transition from $A$ to $B$ contributes the likelihood ratio $1/2$, and each transition from $B$ to $A$ contributes the likelihood ratio $2$. Hence, the product of likelihood ratios around the loop is
\[
    \prod_j \frac{\tau_2(s\mid\omega_j)}{\tau_2(s\mid\overline\omega_j)}
    =2^{\#(B\to A)-\#(A\to B)},
\]
which is not equal to one whenever the loop is not balanced.

By contrast, any $F_1$-measurable experiment $\tau_1$ must satisfy
$$
    \prod_j \frac{\tau_1(s\mid\omega_j)}{\tau_1(s\mid\overline\omega_j)}=1,
$$
because $\overline\omega_j$ and $\omega_{j+1}$ are in the same $F_1$-atom.  Thus Oracle~1 cannot reproduce this likelihood vector signal by signal; Lemma~\ref{Lemma - cover necessity} establishes the corresponding obstruction to posterior-law mimicry while allowing signal relabeling and splitting. This is the likelihood-ratio obstruction behind Proposition~\ref{proposition: balanced}: for irreducible loops, balancedness is exactly the algebraic condition that rules out this obstruction, and the proposition shows that it is equivalent to the geometric condition of being \(F_2\)-covered.

\end{document}